\documentclass[iop,numberedappendix]{emulateapj}
\usepackage{amssymb,amsmath}
\pdfoutput=1
\newcommand{\HST}{\emph{HST}}
\newcommand{\Subaru}{\emph{Subaru}}
\newcommand{\Spitzer}{\emph{Spitzer}~}
\newcommand{\Hauto}{H_{\small{\rm auto}}}
\newcommand{\msol}{M_{\odot}}

\newcommand{\SExtractor}{{\tt SExtractor}~}
\newcommand{\Sersic}{S\'{e}rsic~}
\newcommand{\SSFR}{\textrm{SSFR}}

\newcommand{\fpair}{f_{\textrm{pair}}}

\newcommand{\code}[1]{\texttt{#1}}

\shorttitle{}
\submitted{Accepted to ApJ}
\begin{document}

\shortauthors{Newman et al.}
\shorttitle{Can minor merging account for the size growth of quiescent galaxies?}

\title{Can Minor Merging Account for the Size Growth of Quiescent Galaxies?\\
New Results from the CANDELS Survey}
\author{Andrew B. Newman${}^1$, Richard S. Ellis${}^1$, Kevin Bundy${}^2$, and Tommaso Treu${}^3$}
\affil{${}^1$ Cahill Center for Astronomy \& Astrophysics, California Institute of Technology, MS 249-17, Pasadena, CA 91125, USA \\
${}^2$ Institute for the Physics and Mathematics of the Universe, University of Tokyo, Kashiwa 277-8582, Japan \\
${}^3$ Department of Physics, University of California, Santa Barbara, CA 93106, USA}
\email{anewman@astro.caltech.edu}

\begin{abstract}
The presence of extremely compact galaxies at $z\sim2$ and their subsequent growth in physical size has been the cause of much puzzlement. We revisit the question using deep infrared Wide Field Camera 3 data to probe the rest-frame optical structure of 935 galaxies selected with $0.4<z<2.5$ and stellar masses $M_*>10^{10.7} \msol$ in the UKIRT Ultra Deep Survey and GOODS-South fields of the CANDELS survey. At each redshift, the most compact sources are those with little or no star formation, and the mean size of these systems at fixed stellar mass grows by a factor of $3.5\pm0.3$ over this redshift interval. The data are sufficiently deep to identify companions to these hosts whose stellar masses are ten times smaller. By searching for these around 404 quiescent hosts within a physical annulus $10~h^{-1}~\textrm{kpc} < R < 30~h^{-1}$~kpc, we estimate the minor merger rate over $0.4<z<2$. We find that $13\%-18\%$ of quiescent hosts have likely physical companions with stellar mass ratios of 0.1 or greater. Mergers of these companions will typically increase the host mass by $6\%\pm2\%$ per merger timescale. We estimate the minimum growth rate necessary to explain the declining abundance of compact galaxies. Using a simple model motivated by recent numerical simulations, we then assess whether mergers of the faint companions with their hosts are sufficient to explain this minimal rate. We find that mergers may explain most of the size evolution observed at $z \lesssim 1$ if a relatively short merger timescale is assumed, but the rapid growth seen at higher redshift likely requires additional physical processes.
\end{abstract}

\section{Introduction}

The compact nature of massive quiescent galaxies at redshifts $z \simeq 2$ was a surprising discovery when it was announced some years ago \citep[e.g.,][]{Daddi05,Trujillo06,Buitrago08,vanDokkum08}. Many red galaxies with stellar masses $M_* \simeq 10^{11} \msol$ have effective radii $R_e \simeq 1$~kpc, $3-5$ times smaller than comparably massive early-type galaxies in the local universe. This suggests that they grew significantly in size, but much less in stellar mass. Initially there was some suspicion that the stellar masses of the $z\simeq2$ sources were overestimated, but deep spectroscopic data \citep{Cappellari09,Newman10,vandeSande11} have verified dynamically the high masses of selected $1<z<2$ sources and, in conjunction with the abundance of dynamical masses for lower redshift sources \citep{Treu05,vanderWel05}, provided a valuable, independent confirmation of the size evolution.

Only two physical explanations have been put forward to explain this remarkable growth in size while avoiding the overproduction of present-day high-mass galaxies. Adiabatic expansion through significant mass loss can lead to size growth \citep{Fan08,Fan10}. A galaxy that loses mass as a result of winds driven by an active nucleus or supernovae, for example, will adjust its size in response to the shallower central potential. However, the ``puffing up'' arising from baryonic mass loss occurs only when the system is highly active and young in terms of its stellar population (\citealt{Ragone-Figueroa11}, see also \citealt{Bezanson09}), so it is difficult to see how this mechanism can account for the gradual and persistent growth in size observed for compact sources that are mostly quiescent in nature.

In a hierarchical picture of galaxy formation, mergers are expected to lead to growth in size and stellar mass. Whereas major mergers, involving nearly equal-mass components, will lead to comparable growth in both size and mass, minor mergers involving lower-mass companions can produce more efficient size growth \citep{Bezanson09,Naab09,Hopkins10}.  This mechanism requires  a high rate of occurrence of minor mergers, a significant fraction of which must involve gas-poor companions. Although the major merger rate is observationally constrained reasonably well over $0<z<1$ \citep[e.g.,][]{Kartaltepe07,Lin08,Bundy09,deRavel09,Lotz11} and via a few measurements up to $z \simeq 3$ \citep[e.g.,][]{Bluck09,Man11}, the rate at which minor merging occurs requires exquisitely deep photometric data. For this hypothesis, the key question is whether observations confirm that minor merging occurs at the required rate.

The infrared Wide Field Camera 3 (WFC3/IR) on board the Hubble Space Telescope (\emph{HST})  enables us to address the question of whether minor merging is sufficiently frequent to account for the size growth of compact sources since $z \simeq 2$. The CANDELS survey (GO 12444/5; PIs: H.~C.~Ferguson and S.~M.~Faber) provides an excellent resource for addressing this question since, in the first two fields to be observed -- the UKIRT Ultra Deep Survey and southern GOODS fields -- the associated ground- and space-based and photometry spanning $0.4-8\mu$m is sufficiently deep not only to identify possible companions ten times less massive than their hosts, but also to reliably determine their photometric redshifts so that a physical association can be evaluated.

Our goal in this paper is thus twofold. First, exploiting the unique combination of depth and angular resolution in the CANDELS near-infrared data, we aim to measure the size growth of massive galaxies. We will show that the most compact sources virtually always have quiescent stellar populations. We then estimate the minor merger fraction by searching for low-mass companions around these quiescent sources within a fixed search annulus of $10~h^{-1}~\textrm{kpc} < R < 30~h^{-1}$~kpc. A physical association can be made through their photometric redshifts. We will then interpret the minor merger fraction as a possible cause for the growth rate of compact massive galaxies.

The plan of this paper is as follows. In Section 2, we introduce the CANDELS WFC3/IR images and the associated photometric data. We describe the selection of 935 galaxies with stellar masses $>10^{10.7}\,M_{\odot}$ in the photometric redshift range $0.4<z<2.5$. Section 3 analyzes the size growth for this sample and compares our results to earlier work. Section 4 introduces our search for faint companions around 404 quiescent galaxies spanning the redshift range $0.4<z<2$ in which we can confidently detect companions with 10\% of the stellar mass of their hosts. We discuss the robustness of our search, make corrections for spurious unassociated pairs, and assess the stellar mass content and colors of these companions. Finally, in Section 5 we interpret our minor merger rate in the context of size growth. After discussing the size growth of the quiescent population, we turn to a test that asks whether the merger rate is consistent with the increasing rarity of compact examples at later times. Finally, we summarize our conclusions and the remaining uncertainties in Section 6.

Throughout the paper, we adopt a concordance cosmology with $(\Omega_m, \Omega_v, h) = (0.3, 0.7, 0.7)$ and use the AB magnitude system \citep{Oke83}.

\section{Data and Catalogs}
\label{sec:data}

We have compiled an extensive database of optical and infrared observations from 
space and the ground in the UKIRT Deep Survey (UDS, \citealt{Lawrence07}) and GOODS-South \citep{Giavalisco04} fields, offering the
wide spectral coverage from 0.4 to $8\mu$m necessary to secure quality photometric redshifts,
stellar masses, and stellar population parameters for mass-complete samples of galaxies to $z \simeq 2.5$. Although our supplementary photometry covers a much wider area, we restrict our attention 
to the CANDELS WFC3/IR footprints, since our program requires the depth and angular resolution 
in the rest-frame optical afforded by \HST.

\subsection{Imaging Data}

The UDS and GOODS-S fields have been observed with \HST/WFC3 in the $J$ (F125W) and $H$ (F160W) filters \citep{Grogin11,Koekemoer11}. 
In the UDS, the v0.5 mosaics of the two epochs of WFC3/IR imaging were coadded.  For the Advanced Camera for Surveys (ACS) F606W 
and F814W imaging in the UDS, we used only the second epoch of observation, since the 
first epoch contained some reduction artifacts at this time of this work. The \HST~imaging was supplemented by 
deep \Subaru~$BVRiz$ imaging from the \Subaru/\emph{XMM-Newton} Deep Survey (SXDS; \citealt{Furusawa08}), using the mosaics prepared by \citet{Cirasuolo10}, and by $K$-band imaging from the UKIDSS UDS Data Release 6 (DR6). 
Deep \Spitzer Infrared Array Camera (IRAC) data from the SpUDS survey (PI: J.~S.~Dunlop) allows us to access the rest-frame near-infrared to $z \simeq 3$. We cross-referenced our catalogs to the SpUDS MIPS catalog using a positional tolerance of 1$''$.

In GOODS-S, we use the first three epochs of WFC3/IR imaging in the CANDELS Deep area and the first 
epoch of the Wide region. To this we add the GOODS $BViz$ ACS imaging, as well as ground-based data 
in $U$, $R$ and $K$ from VIMOS \citep{Nonino09} and ISAAC \citep{Retzlaff10} at the Very Large Telescope (VLT). The two epochs of ultradeep IRAC imaging from the \Spitzer GOODS Legacy Science Program (PI: M.~Dickinson) were co-added to produce a single mosaic. We again cross-referenced our catalog to the MIPS catalog.

\subsection{Catalogs\label{sec:catalogs}}

For the main photometric catalog, we chose the WFC3 $H$ band as the detection image, thereby taking 
advantage of the high-resolution \HST~imaging while maintaining a selection that is as complete in stellar mass as possible. The $H$ mosaic, distributed on a 60~mas pixel scale, was rebinned to a 120~mas scale, and all other imaging was registered to this grid. Object detection and photometric precision are insignificantly affected by this slightly coarser sampling, but the computational efficiency is greatly increased. For measurement of structural parameters, where the highest possible resolution is critical, we created catalogs for each \HST~mosaic at the original scale (60 mas for WFC3/IR and 30 mas for ACS) and matched these to the main catalog.

Each image (ground, \HST, and IRAC) was first registered to the $H$-band mosaic using smooth transformations  as determined by the IRAF task {\tt geomap}. The images were then drizzled onto the uniform grid, precisely  conserving flux, using {\tt geotran}. A composite point spread function (PSF) was constructed in each image by stacking suitably normalized cutouts of bright, unresolved sources. Matching PSFs is critical for accurate colors  across images of widely varying resolution, yet one wishes to avoid unnecessary degradation of the high-resolution  data as far as possible. We struck the following compromise: the ACS and WFC3 $J$ images, each of higher resolution  than the detection $H$ image, were convolved to match the $H$-band PSF. Colors were then measured in fixed  apertures of $1\farcs5$ diameter by running \SExtractor \citep{Bertin96} in dual image mode. For the lower-resolution imaging from  ground-based instruments and IRAC, we measured the $X$ band flux $f_X$ in a wider aperture (see below) appropriate to the PSF in a given band $X$. We then convolved the $H$ image to match the $X$ PSF and measured the $H$ flux $f_{H,{\textrm{wide}}}$ in the wide aperture. Finally, $f_X$ was scaled by the ratio $f_{H,1\farcs5} / f_{H,{\textrm{wide}}}$,  in order to refer all fluxes to a common aperture. In this way, the \HST~resolution is degraded as minimally as necessary for each band.

To determine a convolution kernel that matches two PSFs, we took the analytic Moffat kernel that best matched the  curves of growth, weighting toward the radii relevant for our aperture photometry. This method typically matched curves of growth to $\simeq 1-2\%$. Colors between \HST~filters were measured in fixed apertures of $1\farcs5$ diameter. For broader PSFs, the aperture diameter was set proportionally to the size of the PSF: $4\times$ the half-light radius, but restricted to lie within the range $1\farcs5 - 3''$. The upper limit was chosen to avoid excessive confusion in the IRAC data. Aperture colors were scaled to total fluxes using the \SExtractor AUTO aperture in the $H$-band image. Photometric uncertainties were determined using apertures placed at random in blank sections of the images. Systematic uncertainties  of 4\% (10\% in the IRAC bands) were added in quadrature to account for zeropoint errors, aperture mismatch, and  color-dependent
flat-field errors in IRAC. Small Galactic extinction corrections were made based on the dust maps of \citet{Schlegel98}.

\subsection{Photometric redshifts and other derived parameters\label{sec:masses}}

Using this photometry spanning $0.4 - 8 \mu$m, photometric redshifts were estimated using the \code{EAZY} code \citep{Brammer08}. We permitted linear combinations of all templates in its default set and adopted the prior based on $K$-band flux. Spectroscopic redshift surveys have been conducted with the VLT in GOODS-S by \citet{Vanzella08}, \citet{Popesso09}, and \citet{Balestra10}, while \citet{Wuyts08} have compiled redshifts from a number of additional sources. In the UDS\footnote{\url{http://www.nottingham.ac.uk/astronomy/UDS/data/dr3.html}} we draw from C.~Simpson et al.~(in preparation), M.~Akiyama et al.~(in preparation), and \citet{Smail08}. Only spectra with high quality flags were included. These spectroscopic data provide an opportunity to test the accuracy of our photometry by forcing \code{EAZY} to fit templates at the known redshifts and averaging the residuals in each filter \citep[e.g.,][]{Capak07}. The resulting systematic offsets were small (typically $\lesssim 0.03$~mag), confirming the high quality of the photometric calibration and PSF matching. The one exception was the VIMOS $R$-band image, to which we added a $-0.10$~mag correction. In Section \ref{sec:photoz}, we assess the accuracy of our photometric redshifts by comparing to this spectroscopic database.

Stellar population parameters, including stellar masses, were measured by fitting the latest S.~Charlot \& G.~Bruzual (2007, private communication) models to the broadband photometry using the \code{FAST} code \citep{Kriek09}. A large grid of models with exponentially-declining star-formation histories was created, with redshifts between 0.01 and 7 in steps of $0.01(1+z)$, ages between $t = 10^{7}$ and $10^{10.1}$ years (always less than the age of the universe) in 32 logarithmic steps, star-formation timescales $\tau$ between $\tau= 10^{7}$ and $10^{10}$ years in 31 logarithmic steps, and dust content varying between $A_V = 0$ and 3 in 31 steps. Solar metallicity, the \citet{Calzetti00} extinction law, and a Salpeter initial mass function (IMF) were adopted. We chose the Salpeter IMF because it may be more appropriate for massive galaxies \citep{Treu10,vDC10,Auger10IMF,Newman11,Spiniello11}, but our analysis is insensitive to this choice since we require only relative stellar masses.  Rather than adopting the stellar population parameters of the single best-fitting model, we obtain the mean of each parameter by marginalizing over the likelihood function. Finally, rest-frame colors were computed using the InterRest code \citep{Taylor09}.

\begin{figure*}
\centering
\includegraphics[width=0.9\linewidth]{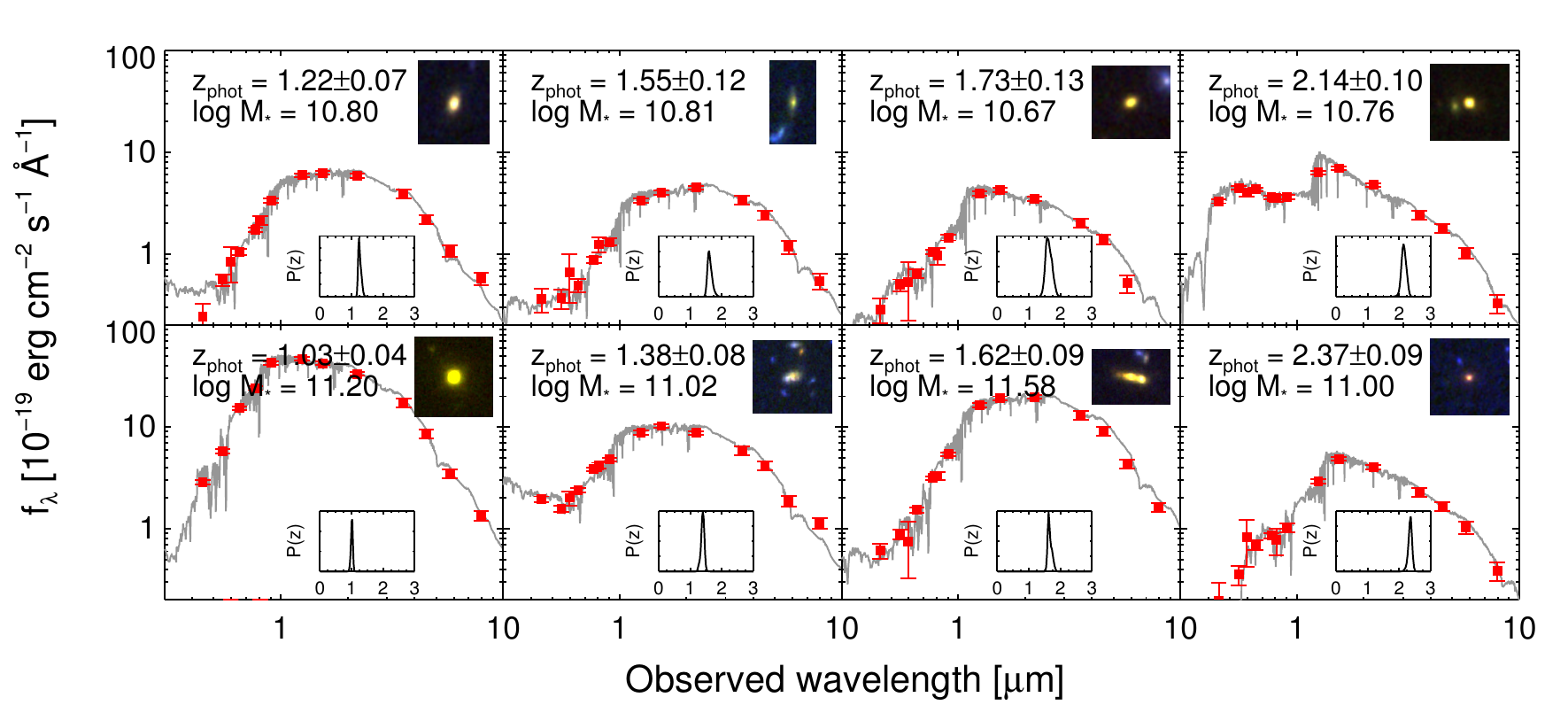}
\caption{Montage of representative massive galaxies at $1 < z <  2.5$. High-precision photometry spanning $0.4-8\mu$m is plotted along with the best-fit spectral synthesis model as described in the text. Composite \HST~images in the $IJH$ filters (where $I$ is F814W or F775W) are inset along with \code{EAZY} photometric redshift 
distributions. \label{fig:montage}}
\end{figure*}

Figure~\ref{fig:montage} displays photometry, spectral energy distribution (SED) fits, redshift constraints, and color composite images for several representative massive galaxies at $1 < z < 2.5$. Note that the signal-to-noise ratio is very high, even at $z \simeq 2$, reflecting the high quality of the photometric data.

\subsection{Survey Mass Limit and Completeness}

We define a limiting stellar mass for our galaxy sample, motivated by the desire to obtain a complete census of satellites with stellar mass ratios $\mu_* = M_{\textrm{sat}}/M_{\textrm{host}} > 0.1$ at $z < 2$ as well as our desire to track evolution in the sizes of mass-selected hosts to $z\simeq2.5$.

\begin{figure}
\centering
\includegraphics[width=1.0\linewidth]{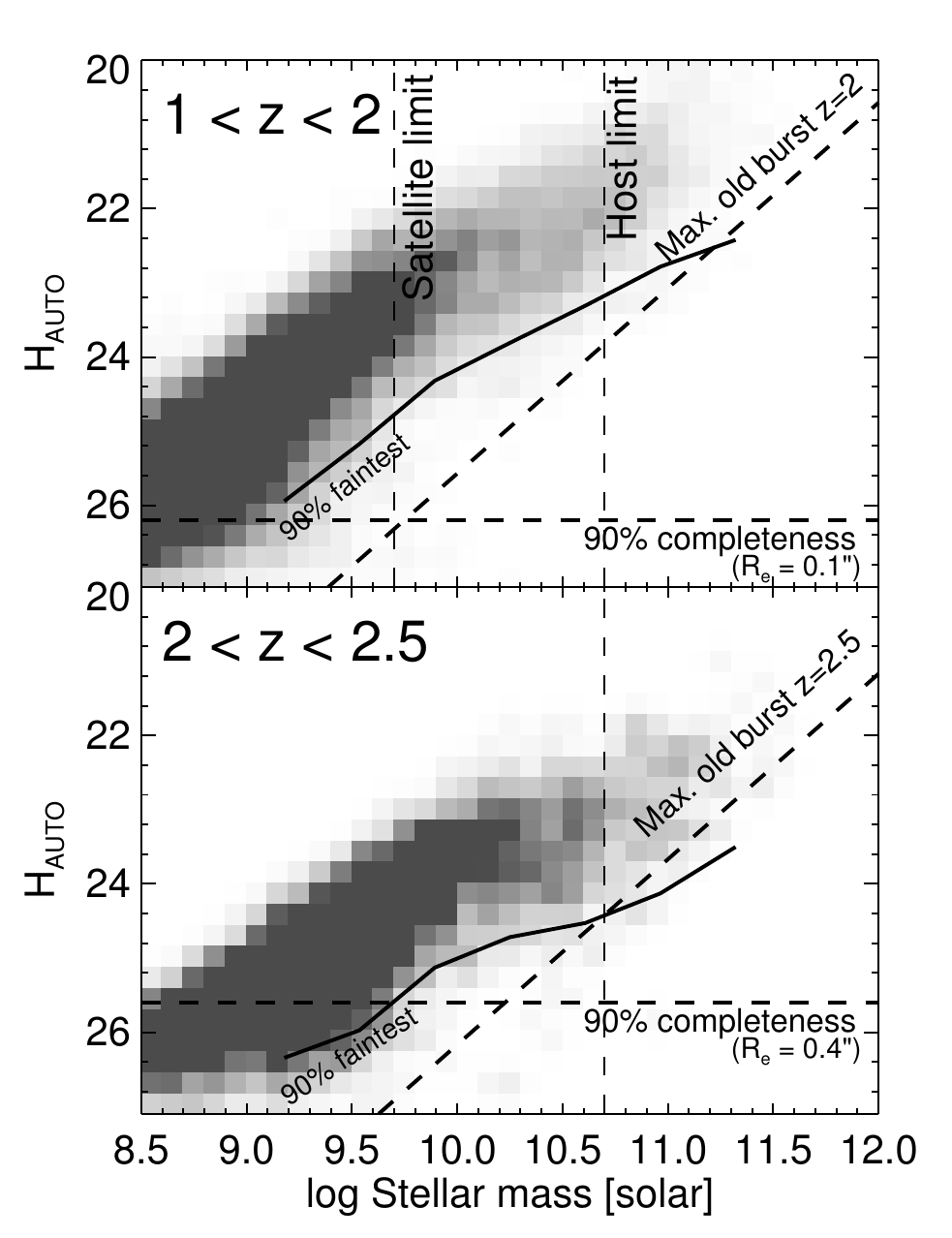}
\caption{Our sample is designed to ensure a complete census of satellites with mass ratios $\mu_* = M_{\textrm{sat}}/M_{\textrm{host} } > 0.1$ at $z < 2$. The relation for a maximally old, dust-free stellar population using the Charlot \& Bruzual (2007) models is shown as a dashed line, while the solid line indicates the 90th percentile in faintness at a given stellar mass. These are compared to completeness levels (horizontal) to set appropriate stellar mass limits. The top panel demonstrates that restricting hosts to $\log M_* > 10.7$ ensures strong detections in $H$ for $\mu_* > 0.1$ satellites at $z < 2$. The bottom panel demonstrates the hosts themselves can be reliably studied to a higher redshift of $z=2.5$.
\label{fig:masslimit}}
\end{figure}

The completeness of our catalog was assessed by inserting synthetic objects into blank sections of the UDS WFC3 $H$ image, blurring by the empirical PSF and binning to the same pixel scale. These were then detected using the same \SExtractor configuration. The 90\% photometric completeness limits are $\Hauto = 26.5$ for point sources and $\Hauto = 25.6, 25.8,$ and 26.1 for de Vaucouleurs profiles with $R_e = 0\farcs4, 0\farcs2,$ and $0\farcs1$, respectively. For de Vaucouleurs profiles with $R_e = 0\farcs1$, which is roughly the size expected for local $\log M_* \simeq 9.7$ early-type galaxies viewed at $z \simeq 2$, the 90\% completeness limit is $\Hauto = 26.1$. Figures~\ref{fig:masslimit}a shows that selecting satellites with $\log M_* > 9.7$ at $z < 2$ ensures $H$-band fluxes above this limit, even for a maximally old population.  Since we demand completeness for $\mu_* > 0.1$, this in turn implies a limit of $\log M_* > 10.7$ for the hosts.

If we are only concerned with studies of the host galaxies, i.e., without the need to detect their faint companions, they can be followed to somewhat higher redshift. We limit ourselves to $z < 2.5$ in order to retain deep detections in F160W, suitable for robust size measurements at our mass limit. Figure~\ref{fig:masslimit}b shows that, in the redshift range $2 < z < 2.5$, we remain complete at $\log M_* > 10.7$ even for $R_e = 0\farcs4$, the most extended profile we tested. This size  corresponds roughly to the size of a local $M_* = 10^{11} \msol$ early-type galaxy viewed at $z=2$. 

\subsection{Surface Photometry and Effective Radii\label{sec:galfit}}

We use \code{Galfit} \citep{Peng10} to fit \Sersic profiles to galaxies in our sample, using an automated procedure to fit adjacent objects simultaneously. The \Sersic index $n$ was restricted to $0.5 < n < 8$, and the size of the fitting box was set by requiring it to enclose the Kron ellipse enlarged by a factor of 2.5. The background was measured in a rectangular annulus extending 40 pixels from the boundary of the fitting box. In order to measure structural parameters at similar rest-frame wavelengths, we selected different filters for fitting according to the redshift. In the UDS, sizes are measured in F814W for $0.4 < z < 0.9$, F125W for $0.9 < z < 1.8$, and F160W for $1.8 < z < 2.5$. In GOODS-S, F775W is used for $0.4 < z < 0.75$, F850LP for $0.75 < z < 1.1$, F125W for $1.1 < z < 1.8$, and F160W for $1.8 < z < 2.5$. This ensures that the wavelength at which sizes are measured always falls in the rest-frame interval $4240 - 6570$~\AA. Based on the mean difference between the \Sersic and AUTO magnitudes in the $H$ band, we applied slight adjustments of $\Delta \log M_* = 0.014n$ to account for light outside of the AUTO aperture. 

An extensive suite of tests performed by randomly inserting synthetic \Sersic profiles into the F814W, $J$, and $H$-band images showed that we are able to recover radii with a typical accuracy of 5-10\%, consistent with other studies \citep{vanderWel08,Newman10}. This procedure automatically incorporates errors arising from background misestimation and blending with neighboring objects, but applies strictly only to symmetric, S\'{e}rsic-like profiles. In the $H$-band image, we additionally tested for possible errors caused by PSF variations by convolving the synthetic profiles with stellar images selected from throughout the mosaic. These were then fitted using the empirical stacked PSF used to analyze the real data. We found that radii as small as $0\farcs05$ (0.4 kpc at $z=2$) can be reliably recovered.

All galaxies with stellar masses exceeding $10^{10.7}\msol$ were fit. For our study of size evolution presented in Section 3, we exclude galaxies for which \HST~imaging in the appropriate filter is not available due to imperfect overlap among the observations (5.7\% of the sample), as well as those whose proximity to the image border or to a bright foreground star or galaxy precluded a reliable measurement (2.8\%). Note that these cuts are uncorrelated with any galaxy property. We also exclude the 5.7\% of remaining galaxies that are fit with a \Sersic index $n  = 0.5$ or 8, i.e., the boundaries of the allowed range of $n$. These size measurements are likely to be unreliable. Although excluding them may slightly bias our mean size measurements, we expect any effect to be minor owing to the small fraction of the sample that they represent.

Effective (half-light) radii are typically reported in a circularized form defined by $R_{e,{\textrm{circ}}} \equiv a \sqrt{q}$, where $a$ is the semi-major axis of the half-light ellipse and $q = b/a$ is the axis ratio. We adopt a slightly different definition: $R_h \equiv a(1+q)/2$. Physically, $R_h$ closely approximates the half-light radius obtained from a classical curve of growth analysis on the intrinsic (PSF-deconvolved) \Sersic profile, i.e., the radius of the circle containing half of the total light, as we verified numerically. This definition differs appreciably from the more common $R_{e,\textrm{circ}}$ only for small $q$, for which the latter diverges from a curve of growth measurement. For our mass-selected sample, the mean (median) difference between $R_h$ and $R_{e,{\textrm{circ}}}$ is only 5\% (2\%) and has no impact on the evolutionary trends that are the main subject of this paper.

\subsection{Comparison to the Sloan Digital Sky Survey}

The total area covered by our UDS and GOODS-S catalogs is 311~arcmin${}^{2}$. At $z < 0.4$, too little volume is probed to provide reasonably large and representative samples of galaxies. In the following analysis, we therefore supplement our catalogs by comparing to $z \sim 0$ galaxies in the Sloan Digital Sky Survey (SDSS DR7; \citealt{SDSSDR7}). We selected galaxies from the spectroscopic survey in the redshift interval $0.05 < z < 0.07$. These were matched to stellar mass and star formation rate estimates from the MPA-JHU DR7 catalog \citep{Kauffmann03}\footnote{\url{http://www.mpa-garching.mpg.de/SDSS/DR7/}} and to \Sersic fits from the NYU Value Added Catalog \citep{Blanton05}. The stellar masses were shifted by $+0.19$~dex to convert from a Kroupa to a Salpeter IMF.

There may be substantial systematic differences between the derived measurements in the SDSS and CANDELS. For example, our SED fits include NIR photometry, while the SDSS does not. Comparisons of effective radii are also uncertain. \citet{Guo09} fit \Sersic profiles to SDSS images of representative massive galaxies. Around $10^{11} \msol$, their effective radii are on average 0.2~dex larger than the \citet{Blanton05} values. Since none of the results in this paper rely on the SDSS data, we simply adopt the MPA-JHU stellar masses and \citet{Blanton05} radii and, where appropriate, we caution how uncertainties in these affect the analysis.

\section{Size Evolution of Massive Galaxies\label{sec:sizeevolution}}

The unique depth, resolution, and area of the CANDELS near-infrared images provides an opportunity to freshly examine the rate of size growth for various categories of galaxies within our mass-selected sample over $0.4<z<2.5$. Below we will focus on evolution in the stellar mass -- size plane:
\begin{equation}
R_h = \gamma \left(\frac{M_*}{10^{11} \msol}\right)^{\beta} = \gamma M_{11}^{\beta}
\label{eqn:masssize}
\end{equation}
In the nomenclature of early-type galaxies, this is the Kormendy projection of the stellar mass fundamental plane (relating $M_*$, $R_h$, and $\sigma$; e.g., \citealt{Auger10}). It has been extensively studied, particularly at high redshift where it is the most observationally accessible projection \citep[e.g.,][]{Trujillo06,Toft07,Trujillo07,Zirm07,Buitrago08,Cimatti08,vanDokkum08,vanderWel08,Damjanov09,Toft09,Mancini10,Ryan10,Saracco11,Damjanov11}. The mass-size plane provides some of the most powerful constraints on the merger histories of galaxies \citep[e.g.,][]{Nipoti03}, which we exploit in Section 5.

\begin{figure*}
\centering
\includegraphics[width=\linewidth]{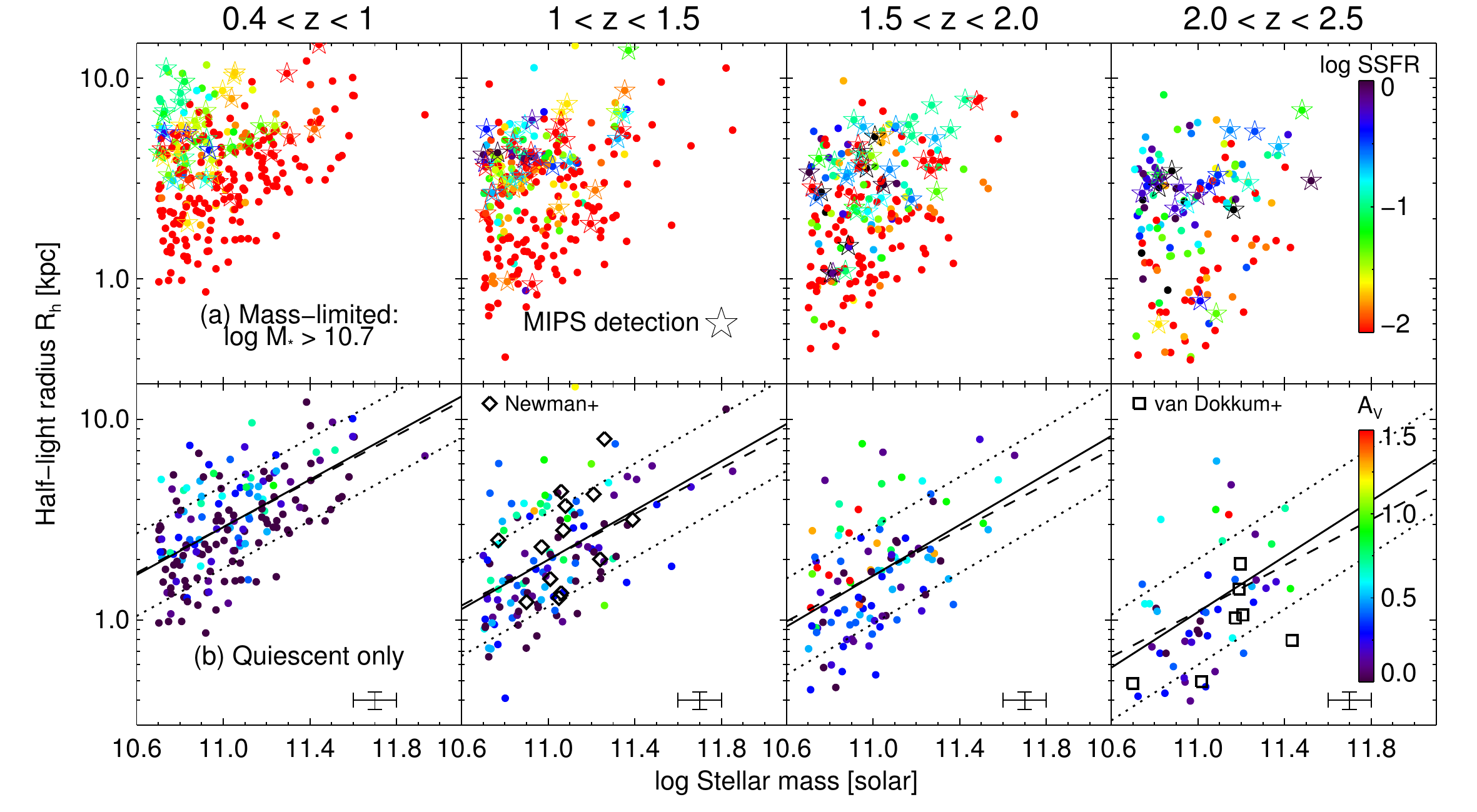}
\caption{Size evolution of massive galaxies over $0.4 < z < 2.5$. \textbf{(a)} All galaxies with 
$\log M_* > 10.7$, with color encoding the SSFR.  At each redshift there is a strong relationship between 
SSFR and size, with the most quiescent galaxies being the most compact. \textbf{(b)} The quiescent subsample, with color now encoding the extinction $A_V$. Linear fits show the best fit to $R_h \propto M_*^{\beta}$  with $\beta$ as a free parameter (solid line) or fixed to the slope $\beta = 0.57$ (dashed). Dotted lines indicate the $1\sigma$ vertical scatter. Spectroscopic samples from \citet{Newman10} and \citet{vanDokkum08} (using CB07 fits from \citealt{Muzzin09}) that pass our selection criteria are plotted as diamonds and squares. Sizes represent \Sersic effective radii measured at rest-frame $\sim5000$~\AA~as described in Section~2.\label{fig:sizeevolution}}
\end{figure*}

Our sample contains 935 galaxies in the interval $0.4 < z < 2.5$ with stellar masses exceeding $\log M_* = 10.7$. Figure~\ref{fig:sizeevolution}a demonstrates a strong correlation between size and the specific star-formation rate (SSFR, the star formation rate per unit stellar mass), such that the most compact galaxies are the most quiescent. The lower envelope of points delineates an evolving  ``compactness'' limit. This figure confirms the results of many previous studies \citep[e.g.,][]{Trujillo06,Franx08,Williams10,Weinzirl11} but represents an important advance, since it is based on a large, homogeneous sample with space-based sizes uniformly measured in the rest-frame optical to $z = 2.5$. The advantage of space-based imaging is particularly evident for lower-mass galaxies with $\log M_* < 11$. Most of these that are quiescent at $z \gtrsim 1.4$ have radii comparable to or smaller than $0\farcs1-0\farcs2$, which is generally taken as the limit for reliable size measurements in seeing-limited data \citep{Bezanson11,Williams11}.

We expect the appearance of the mass-size plane to change with time both through the evolution of existing galaxies and the continued emergence of new systems \citep[e.g.,][]{Robertson06,Hopkins10b}. Nevertheless, the evolution of the compactness threshold is strong enough that by $z \sim 2.5$, the most compact galaxies are typically smaller than any galaxy found in the lowest redshift bin. Although there may be a few compact systems persisting even to $z = 0$ \citep{Valentinuzzi10}, their comoving number density is clearly greatly depleted \citep{Trujillo09,Taylor10}. This implies that individual, compact high-$z$ systems must grow in size, and that the responsible processes must evacuate the most compact regions of the mass-size plane at a rate consistent with Figure~\ref{fig:sizeevolution}. For this reason, in the following we concentrate foremost on quiescent galaxies, which are the most compact.

Figure~\ref{fig:sizeevolution}b shows the trends we find for 483 quiescent galaxies, defined as the subsample with 
$\SSFR < 0.02~\textrm{Gyr}^{-1}$ and no detection in the MIPS $24\mu$m channel, which would indicate the presence of warm dust. Several other definitions of quiescence are common in the literature. Among these, we note that 88\% of our quiescent sample would be selected by the $UVJ$ color cuts introduced by \citet{Williams10}. The median \Sersic index of the quiescent subsample evolves modestly, from $\langle n \rangle \simeq 3$ to 4.5 over our entire redshift baseline, while the median axis ratio is essentially constant at $\langle q \rangle = 0.66$. This is consistent with the majority of these galaxies being bulge-dominated, although some are surely disks \citep[see][]{Kriek09b,vanderWel11}. 

Solid lines in Figure~\ref{fig:sizeevolution} show fits to Equation \ref{eqn:masssize}, which are reported in Table~\ref{tab:sizeevolution}. Interestingly, there appears to be little or no evolution in the slope $\beta$ of the mass--radius relation within the present uncertainties: formally, we find $d\beta/dz = 0.05\pm0.10$. Further, the mean $\langle \beta \rangle = 0.61 \pm 0.05$ is consistent with the $\beta = 0.57$ we measure for galaxies selected in the SDSS using the same stellar mass and SSFR criteria. In the context of spheroids, it is known that this slope cannot be established solely by dry mergers of smaller systems \citep[e.g.,][]{Ciotti07}, and that it must therefore be imprinted by dissipational processes during a spheroid's formation, i.e., before it becomes quiescent. From this perspective, it is perhaps expected that the mass--radius slope for quiescent systems should persist to very early epochs.

Fits to the mass-size relation are always subject to an Eddington bias arising from the steep mass function. This steepness implies that near the limiting mass threshold, lower-mass galaxies are scattered above the threshold more frequently than higher-mass galaxies are scattered below it. We estimated this bias through Monte Carlo simulations, generating mock data with errors in stellar masses and radii typical of our sample. These were fit to a linear relation using a simple least-squares regression with equal weighting, as was done for the real data. The measured $\beta$ may underestimate the true slope by $0.02-0.05$. Since this correction is small, sensitive to the true errors in the stellar mass estimates, and similar at each redshift, we decided not to apply it.

\begin{deluxetable}{cccc}
\tablewidth{\linewidth}
\tablecolumns{4}
\tablecaption{Fits of the Mass-Size Relation of Quiescent Galaxies to $\log R_h = \gamma + \beta (\log M_* - 11)$}
\tablehead{\colhead{Redshift} & \colhead{$\gamma$} & \colhead{$\beta$} & \colhead{$\sigma_{\log R_h}$}}
\tablecomments{Fits are plotted in Figure~\ref{fig:sizeevolution}. Errors are determined from bootstrap resampling (negligible in the SDSS). The observed scatter is measured using the standard deviation.\label{tab:sizeevolution}}
\startdata
SDSS $z=0.06$ & 0.54 & 0.57 & 0.16 \\
$0.4 < z < 1.0$ & $0.46 \pm 0.02$ & $0.59 \pm 0.07$ & $0.21\pm0.01$ \\
$1.0 < z < 1.5$ & $0.30 \pm 0.02$ & $0.62 \pm 0.09$ & $0.23\pm0.02$ \\
$1.5 < z < 2.0$ & $0.21 \pm 0.02$ & $0.63 \pm 0.11$ & $0.24\pm0.02$ \\
$2.0 < z < 2.5$ & $0.04 \pm 0.04$ & $0.69 \pm 0.17$ & $0.26\pm0.03$
\enddata
\end{deluxetable}

\begin{figure*}
\plottwo{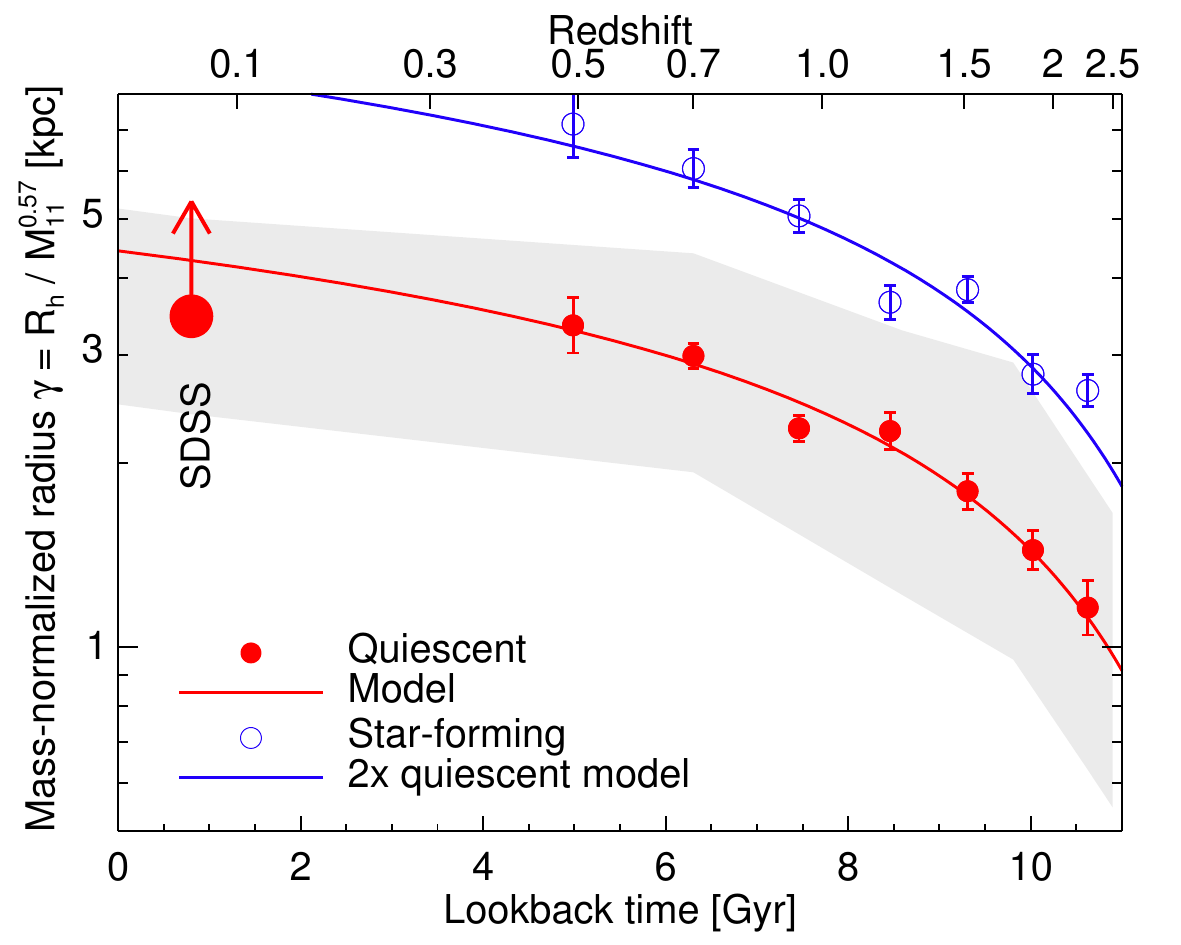}{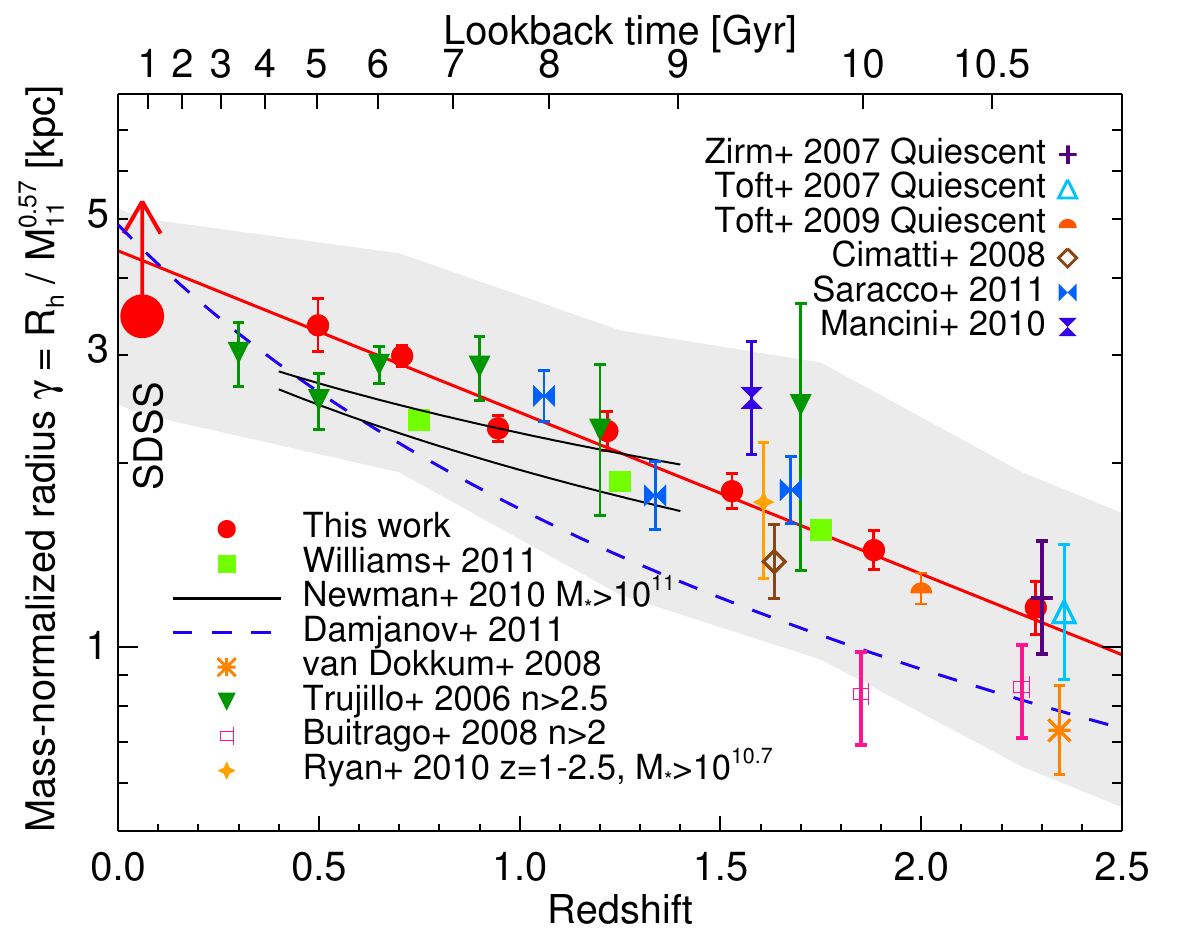}
\caption{\textbf{Left:} Evolution in the mean size of quiescent (red) and star-forming (blue) galaxies, measured at matched rest-wavelength and normalized to $M_* = 10^{11} \msol$ using the slope $R_e \propto M_*^{0.57}$. Error bars indicate the $1\sigma$ uncertainty in the mean, accounting for random sampling errors only. The shaded region shows the $1\sigma$ scatter in the quiescent population as measured in Table 1. The large red dot indicates our default SDSS relation; the arrow estimates the change if the \citet{Guo09} sizes were used instead (Section 2.6). \textbf{Right:} Our results (red circles) are compared to other recent estimates, as indicated by the legend.\label{fig:oned_growth} }
\end{figure*}

Noting the lack of significant evolution in the \emph{slope} of the mass--size relation of quiescent galaxies, we fix $\beta = 0.57$ (the SDSS slope) and consider the growth of the normalization $\gamma$ in Figure~\ref{fig:oned_growth}a. This figure displays the mean size of quiescent systems normalized to a stellar mass of $10^{11} \msol$. It is important to recognize that the figure concerns the size evolution of the population as a whole and not necessarily the growth rate of any individual galaxy. Accordingly, we note that the growth rate at fixed mass $d \log \gamma / dt$ accelerates over this interval, remaining fairly gradual at $z \lesssim 1$ and then noticeably increasing over $z \approx 1-2.5$. We reached the same conclusion in \citet{Newman10}. Figure~\ref{fig:oned_growth}b shows the same data plotted against redshift; there is no apparent change in $d \log \gamma / dz$. We concentrate here on the evolution per unit time because it most directly relates to the effects of mergers. The blue points in Figure~\ref{fig:oned_growth}a indicate the sizes of the star-forming systems in our mass-limited sample. Interestingly, the evolution in size is similar to that for the quiescent galaxies, so that star-forming galaxies are always, on average, a factor of $\simeq 2$ larger than quiescent systems of the same mass over the entire redshift range (see \citealt{Law11}).

Figure~\ref{fig:oned_growth}b compares our results on quiescent galaxies to several recent studies. Overall, there is a fair degree of convergence given the diverse nature of the samples, which apply various selection techniques to different types of data (e.g., sizes measured in different wavebands, from space and the ground, selection by color or morphology). In compiling these data we have harmonized all stellar masses to a Salpeter IMF and have applied an additional correction of $\Delta \log M_* = -0.05z$ for data fit with \citet[][BC03]{Bruzual03} models.\footnote{This accounts for the average difference between BC03 and CB07 stellar mass estimates in our quiescent sample. The redshift dependence is expected, since the TP-AGB phase that distinguishes these models is predominant at ages of $\sim 1$~Gyr.} We caution that direct comparisons of simple parametric fits may be misleading, since these can depend strongly on the redshift interval that is fit.

The primary conclusion from the high-quality CANDELS data now in hand is a factor of $3.5\pm0.3$ growth in size at fixed stellar mass for quiescent sources over the redshift interval $0.4<z<2.5$, with evidence for accelerated growth at earlier times (Figure~\ref{fig:oned_growth}a). Our challenge in the remainder of the paper will be to attempt to explain this growth rate. Although most workers have focused on the growth of the {\it mean size at a given epoch} (Figures 4), there is valuable information in the {\it distribution of sizes} which can be used to discriminate between the growth of individual systems over time and the arrival of new members of the population.  Although we will discuss this model in more detail in Section 5, it is helpful to describe the data in terms of the evolving size distribution at this juncture.

\begin{figure}
\includegraphics[width=\linewidth]{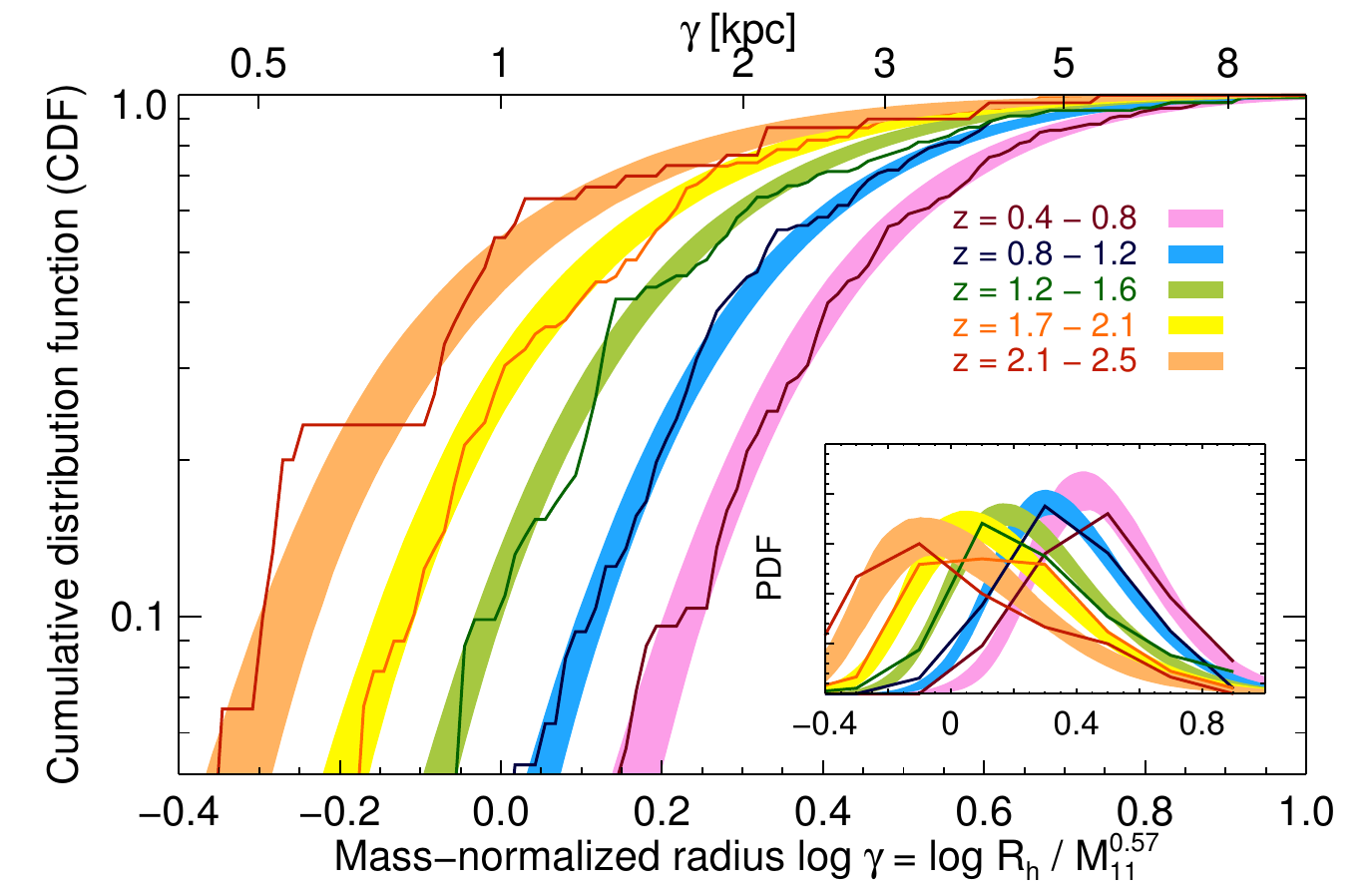}
\caption{Cumulative distribution of the mass-normalized radius $\gamma = R_h / M_{11}^{0.57}$ in several redshift bins. The model described in the text is overlaid in bands with widths indicating the 90\% confidence interval. Differential distributions are inset.\label{fig:sizemodel}}
\end{figure}

Figure~\ref{fig:sizemodel} shows the cumulative and differential (inset) distributions of the mass-normalized radius $\gamma$ for quiescent galaxies in several redshift bins. The distribution is positively skewed in the higher redshift bins, i.e., it exhibits an excess of galaxies with large $\gamma$, which is mostly clearly visible in the inset. The largest quiescent galaxies at a given mass frequently show signs of dust (see coloring in Figure~\ref{fig:sizeevolution}b), suggesting that their rest-optical sizes are impacted by central attenuation.\footnote{To illustrate the effect of extinction, if we restrict to the $\sim80\%$ of quiescent galaxies with $A_V < 0.6$, the intercepts in Table 1 decrease by $\Delta \log \gamma = -0.05$ at $z > 1$, the slopes vary by $<1\sigma$, and the scatter becomes $\sigma_{\log \gamma} = 0.20$~dex in every redshift bin.} For our study in Section 5, the driving quantity is the declining abundance of compact galaxies. Therefore, when fitting the size distributions, it is important to adopt an asymmetric form so that the distribution at small $\gamma$ is not affected by a few apparently large galaxies.

With this in mind, we describe the size distribution at a given redshift with a model in which $\log \gamma$ follows a skew normal distribution. The skew normal distribution has three parameters: the mean $\langle \log \gamma \rangle$, the standard deviation $\sigma_{\log \gamma}$, and a shape parameter $s$ that is related to the skewness. Appendix A summarizes the relevant mathematical details. We parameterize the evolution in each parameter as linear in redshift:
\begin{equation}
\langle \log \gamma \rangle(z) = \langle \log \gamma \rangle_{z=1} + \frac{d \langle \log \gamma \rangle}{dz}(z-1),
\end{equation}
and similarly for $\sigma_{\log \gamma}$ and $s$. We then used a Markov Chain Monte Carlo procedure to sample the likelihood function. Each galaxy was weighted inversely to the number of galaxies at similar redshift to ensure that the entire redshift range contributed equally to the fit. Figure~\ref{fig:sizemodel} compares the observed distribution in $\log \gamma$ to the model with parameters listed in Table~\ref{tab:sizemodel}.

\begin{deluxetable}{ll}
\tablecolumns{2}
\tablewidth{\linewidth}
\tablecaption{Size Evolution Model\label{tab:sizemodel}}
\tablecomments{Mean quantities, marginalized over all other parameters, are reported along with their $1\sigma$ uncertainty.}
\startdata
\hline Mean: $\langle \log \gamma \rangle(z=1)$ & $0.38 \pm 0.01$ \\
$d \langle \log \gamma \rangle / dz$ & $-0.26 \pm 0.02$ \\
Standard deviation: $\sigma_{\log \gamma}(z=1)$ & $0.22 \pm 0.01$ \\
$d \sigma_{\log \gamma} / dz$ & $0.044 \pm 0.017$ \\
Shape: $s(z=1)$ & $2.3 \pm 0.4$ \\
$ds/dz$ & $1.0 \pm 1.1$
\enddata
\end{deluxetable}

This simple model accurately captures the observed features of the size evolution. First, $\langle\log \gamma\rangle$ evolves nearly linearly in redshift as $-0.26z$, which Figures~\ref{fig:oned_growth}b and \ref{fig:sizemodel} demonstrate is a good fit. Second, the scatter $\sigma_{\log \gamma}$ evolves fairly little with redshift. The mild increase is driven mostly by the increasing abundance of large, dusty systems toward higher redshifts, as discussed previously (see footnote 4). Note that we have not attempted to deconvolve errors arising from uncertainties in the stellar masses and radii of individual galaxies. Assuming the formal stellar mass uncertainties and a $10\%$ uncertainty in the radii, the error in individual $\log \gamma$ measurements would be 0.07 dex nearly independent of redshift. Since this is much smaller than the measured width of the distribution, the intrinsic widths would be only $\sim 0.01$~dex smaller than the measured ones. If the true errors were instead twice these estimates, the intrinsic widths would be $\sim 0.05$~dex smaller than the measured ones. The impact of measurement errors is developed further in Appendix B.

\section{Satellites of Quiescent Galaxies at $0.4 < z < 2$\label{sec:satellites}}

The most frequently invoked and well-motivated physical process behind the strong, regular size evolution presented in Section~3 is merging \citep[e.g.,][]{vanderWel08,Bezanson09,Naab09,Hopkins10b}. Most previous studies of merger rates have been confined to $z \lesssim 1.4$ or have focused on ``major'' mergers with stellar mass ratios $\mu_* \gtrsim 0.25$. This is partly due to observational limitations, since probing higher redshifts and lower-mass companions requires deep near-infrared data, and also because major mergers are of special interest for studies tracking morphological transformations.

Size growth, as well as spheroid formation \citep{Bundy07}, is unlikely to be explained by major merging alone. Major mergers are rare: \citet{Bundy09} estimate a rate of only $0.03-0.08~\textrm{Gyr}^{-1}$ for $>10^{10.5} \msol$ galaxies over $0.4 < z < 1.4$. If such low rates persist to $z = 2$, then $\lesssim 15\%$ of galaxies present at $z = 2$ will experience \emph{any} major mergers by $z = 1$, whereas substantial size growth must occur over the same period. ``Minor'' mergers involving lower mass ratios may be crucial. 

\begin{figure}
\includegraphics[width=1\linewidth]{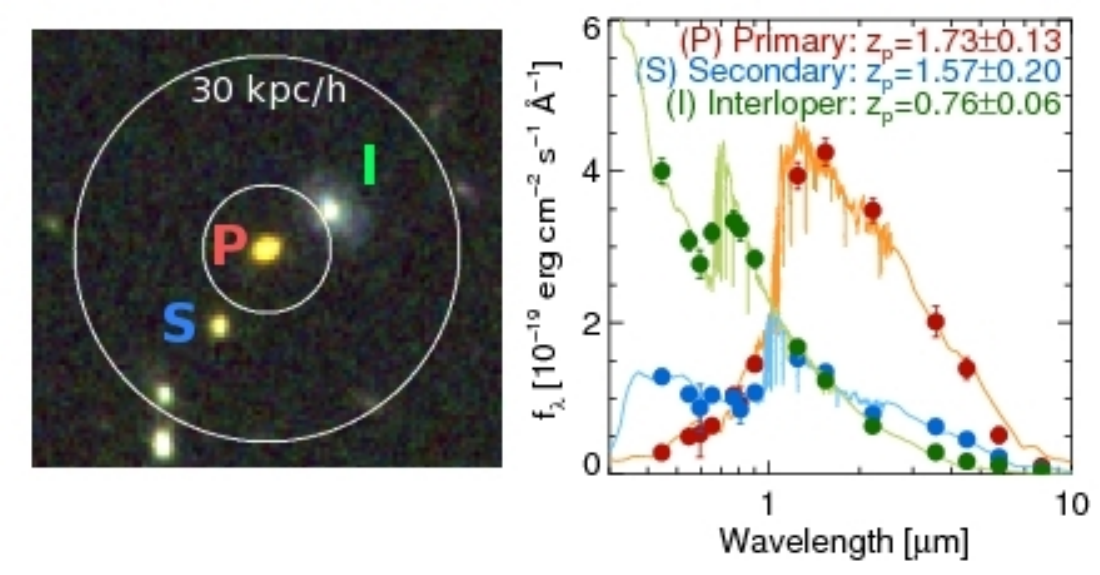}
\caption{Demonstration of the pair counting procedure. The left panel displays a composite F160W/F125W/F814W image
around a $10^{10.7} \msol$ quiescent ``primary'' galaxy at $z_p = 1.73$. The $10~h^{-1}~\textrm{kpc} < R < 30~h^{-1}$~kpc
search annulus is outlined. One $\mu_* \simeq 1:8$ secondary ``S'' is identified as a possible physical association based on its consistent photometric redshift (right panel). A blue galaxy ``I''  within the search aperture is excluded based on its low photometric redshift. The right panel shows the SEDs and best-fitting \code{FAST} models. For clarity, the models have been smoothed and the  fluxes of the interloper reduced by a factor of 2.5.\label{fig:examplesat}}
\end{figure}

In this section, we measure the incidence of close companions to the same set of massive, quiescent galaxies at $0.4 < z < 2$ whose rate of growth was charted in Section 3. As discussed in Section 2, we here limit ourselves to $z < 2$ in order to maintain completeness for stellar mass ratios $\mu_* > 0.1$. Below, we refer to this quiescent sample as the \emph{primary} sample, while the population of potential satellites is called the \emph{secondary} sample. We search for secondaries around each primary galaxy at projected separations of $10~h^{-1} < R < 30~h^{-1}$~proper kpc with stellar mass ratios $0.1 < \mu_* < 1$. Note that the upper limit avoids double counting. In principle, the size of the search annulus should not matter for measuring merger rates, since the merger timescales increase with the search area. In practice, the inner radius avoids searching for secondaries buried within the light of the primary at low redshift, while the outer radius strikes a reasonable balance of finding useful numbers of pairs without being dominated by chance alignments.

Since many galaxies that are close in projection lie at different redshifts, we attempt to secure physical associations by additionally requiring that secondaries have a photometric redshift consistent with the primary, as detailed below. An example of this is given in Figure~\ref{fig:examplesat}.
However, due to the coarseness of photometric redshift estimates, some galaxies selected by this method will still be chance alignments not physically associated with the primary. This contamination rate is estimated simply by randomizing the positions of the primaries throughout the imaging area, maintaining all their other properties, and repeating the search for secondaries using the same criteria. This procedure is repeated many times to improve the statistical accuracy. Below we distinguish \emph{projected} secondaries, which comprise all secondaries found within the search apertures, from the statistical secondary population that remains \emph{after} chance contaminants are correct for as just described, which we term \emph{physical} secondaries. As we discuss in Section 5, it is important to realize that some fraction of these physical secondaries will not be bound to their primary host and therefore only represent candidate satellites or future mergers.

Below we measure the mean number of physical secondaries per primary host and assess the stellar mass content and colors of these systems. To examine the redshift dependence of these quantities, we break the primary sample into three redshift bins of $z=0.4-1$, $z=1-1.5$, and $z=1.5-2$. In Section 5, we turn to the question of whether the size growth measured in Section 3 is consistent with the merger rates inferred here.

\begin{figure}
\centering
\includegraphics[width=0.8\linewidth]{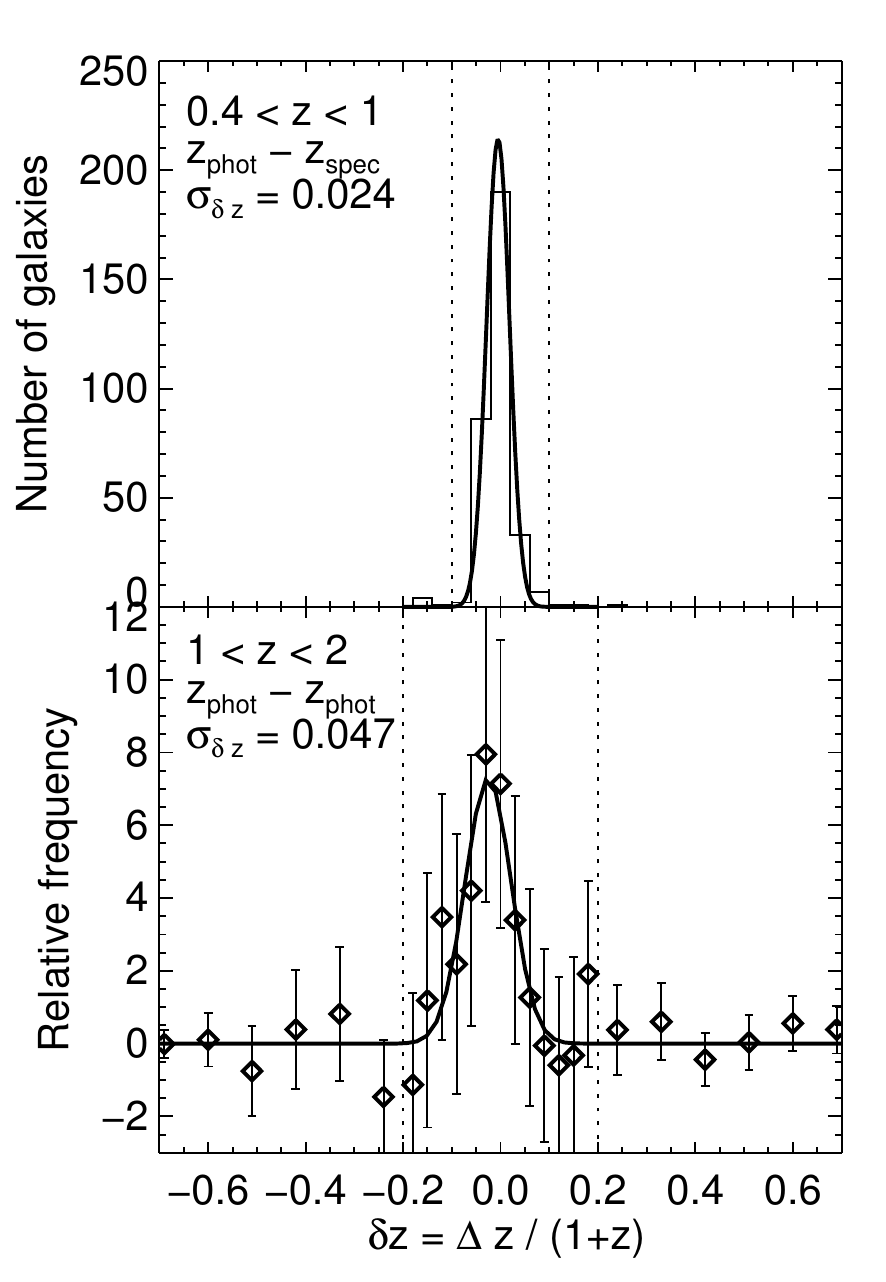}
\caption{Photometric redshift errors in two redshift bins. \textbf{Top:} Comparison with spectroscopic redshifts for $\log M_* > 9.7$ galaxies at $0.4 < z < 1$ indicating a small scatter $\sigma_{\Delta z/(1+z)} = 0.024$. \textbf{Bottom:} Excess probability of a given redshift difference $\Delta z_{\textrm{phot}} / (1+z_1)$ for secondaries within $30~h^{-1}$~kpc of the primary quiescent galaxy sample at $1 < z < 2$. Here secondaries are selected based on $H$ flux ratios as described in the text. The quoted uncertainty refers to the difference in two photometric redshifts. In both panels, the dotted lines indicate the adopted $\delta_z$ threshold. \label{fig:photoz}}
\end{figure}

\subsection{Photometric Redshift Accuracy\label{sec:photoz}}

The secondary galaxy sample is selected to have stellar mass ratios $0.1 < \mu_* = M_2 / M_1 < 1$ and photometric redshift differences $\delta_z = (z_2 - z_1) / (1+z_1)$ less than a fixed threshold, where the subscripts 1 and 2 refer to the primary and secondary galaxies. Determining an appropriate threshold for $\delta_z$ requires knowledge of the accuracy of the photometric redshifts. At $0.4 < z < 1$, 327 of 1244 galaxies with $\log M_* > 9.7$ (the lowest secondary mass we might consider) have spectroscopic redshifts from the sources described in Section 2. Figure \ref{fig:photoz}a compares these to photometric redshifts, demonstrating a small scatter of $\sigma_{\delta z} = 0.024$.\footnote{Throughout, we measure this scatter using the normalized median absolute deviation; see, e.g., \citet{Brammer08}.} We verified that the redshifts, colors, and masses of the spectroscopic subsample at $z < 1$ are reasonably representative of the parent population, so the measured scatter should reflect the true photometric redshift uncertainties. Based on this result, we adopt a threshold of $|\delta_z| < 0.1$. In 2.8\% of cases, the photometric estimates differ ``catastrophically'' by $|\delta_z| > 0.1$.

At $z>1$ the availability of spectroscopic redshifts ($z_{\textrm{spec}}$) declines rapidly. Of the massive, quiescent galaxy sample, 40 galaxies at $z=1-2.3$ have measured $z_{\textrm{spec}}$, of which only 4 are $z > 1.4$. The corresponding photometric redshifts display a small scatter $\sigma_{\delta z} = 0.023$ with only one outlier.\footnote{This single $z_{\textrm{spec}}$ also disagrees with the photometric redshifts in the MUSYC \citep{Cardamone10} and FIREWORKS \citep{Wuyts08} catalogs.} For the full mass-limited sample with $\log M_* > 9.7$ at $z = 1 - 1.5$, the 152 available $z_{\textrm{spec}}$ indicate a scatter of $\sigma_{\delta z} = 0.021$, while at $z = 1.5 - 2$ the 32 available $z_{\textrm{spec}}$ indicate $\sigma_{\delta z} = 0.058$. We have excluded X-ray sources in these comparisons, since they are over-represented in the spectroscopic data. We also note that the vast majority of $z_{\textrm{spec}}$ at $z \gtrsim 1$ are in GOODS-S, so we must assume that similar techniques produce similar results in the UDS. Since $\sigma_{\delta z}$ appears to increase toward $z = 2$, we adopt a wider selection $|\delta_z| < 0.2$ for selecting secondaries at $z = 1 - 2$. With this selection, the catastrophic error rate ($|\delta_z| > 0.2$) is $3 \pm 1$\% and $6\pm4$\% at $z = 1 - 1.5$ and $1.5 - 2$, respectively, based on the available spectroscopic data.

Since the spectroscopic samples are not representative of the full massive galaxy population at $z \gtrsim 1$, it is useful to assess the accuracy of photometric redshifts by other means. We use the empirical technique proposed by \citet{Quadri10}. Their method is an application of the general procedure employed throughout this section: determine the distribution of $\delta_z$ for well-defined primary and secondary samples, and subtract the distribution obtained with scrambled galaxy positions. In this situation, it is preferable to define a secondary sample based on flux rather than stellar mass, since errors in $z_p$ and stellar mass are correlated. To determine a limiting flux ratio that best mimics a mass-based selection $0.1 < \mu_*<1$, we examined the distribution of $\Delta H=H_2-H_1$ between the primary quiescent sample and physical secondaries selected based on their stellar mass. In 90\% of cases, $\Delta H < 2.2$~mag. This motivates a secondary sample defined by $0 < \Delta H < 2.2$~mag. Figure~\ref{fig:photoz}b shows the distribution of redshift differences $\delta_z$ for the physical secondaries.

The distribution is broader than at $z<1$, as the spectroscopic comparison also indicated. The uncertainty $\sigma_{\delta z} = 0.047$ measured here refers to that in the \emph{difference} between two photometric redshifts. The more important uncertainty for this study is the rate of catastrophic ($|\delta_z| > 0.2$) redshift errors. A crude estimate of this can be obtained by integrating the curve in Figure~\ref{fig:photoz}b, which yields $9 \pm 15\%$ over $z = 1 - 2$. Using the same technique at $z \simeq 2$, we find a possibly higher catastrophic rate of $15 \pm 20\%$, but this cannot be determined precisely with the present sample size. These noisier estimates may be higher than the $3-6\%$ inferred from the spectroscopic database, but that sample is biased toward bright systems. A better assessment of the catastrophic rate will require spectroscopic redshifts for larger and more representative samples of galaxies at $z = 1-2$ than is currently available.

\subsection{Subtraction of Host Light}

A concern in all pair studies is that the photometry of the secondary galaxies may be contaminated by light from the hosts.  By inserting synthetic pairs of galaxies with $1:10$ luminosity ratios and projected separations $10-30~h^{-1}$~kpc into the $H$-band mosaic, we found that our detection efficiency is not affected by the proximity of host. Further, these tests indicated that the aperture colors are less affected than the $H$-band AUTO magnitude used to scale the total stellar masses. To correct for this, we measure the \SExtractor AUTO magnitudes of the secondary galaxies in images from which the light of the primary galaxy has been subtracted using our \Sersic fits (Section \ref{sec:galfit}). We also compute stellar mass ratios $\mu_*$ using fits that omit the IRAC photometry, which is the most susceptible to contamination, although this has little effect on our results.

\subsection{Abundance and Stellar Masses of Physical Secondaries}

We now turn to the frequency of physical secondaries and their stellar mass content. First, we consider the pair fraction $\fpair$. This is simply the mean number of physical secondaries per primary galaxy: $\fpair = (N_p - N_r) / N_g$, where $N_p$ is the number of projected secondaries, $N_r$ is the expected number of chance alignments given the total search area, and $N_g$ is the number of primary galaxies. Throughout our pair analysis, we exclude the shallower ``Wide'' section of GOODS-S and primaries for which more than 20\% of the search annulus is masked (e.g., near the image edge). All results in the remainder of this section pertain to mass ratios $0.1 < \mu_* < 1$.

Table~\ref{tab:pairs} presents the results. For quiescent primaries, we find $\fpair = 16\% \pm 3\%$ when averaged over the entire host mass and redshift range. Moreover, the pair fraction does not appear to evolve significantly with redshift within our uncertainties: formally, we find $\fpair \propto (1+z)^{-0.11 \pm 0.68}$. The paradoxical result that the galaxy merger rate remains flat as the halo merger rate increases with redshift has been explored in many theoretical works \citep[e.g.,][]{Berrier06,Kitzbichler08}. Although our present sample is not large enough to be divided in both redshift and mass, we can examine possible mass-dependent trends by dividing the sample into the three mass bins listed in Table~\ref{tab:pairs} and averaging over the full redshift range. We find that the pair fraction increases slightly with stellar mass as $\fpair \propto M_*^{0.28 \pm 0.41}$, in agreement with \citet{Bundy09}. For later use in our models of size growth, we also tabulate the ``intrasample'' fraction $f_{\textrm{IS}}$ of physical secondaries which are also members of the primary sample (i.e., are quiescent and $>10^{10.7} \msol$).

\begin{deluxetable*}{llcccccclc}
\tablewidth{\linewidth}
\tablecolumns{10}
\tablecaption{Abundance and Properties of Physical Secondaries with $0.1 < \mu_* < 1$}
\tablehead{\colhead{Redshift} & \colhead{Primary} & \colhead{$N_p$} & \colhead{$N_r$} & \colhead{$N_g$} &
\colhead{$\fpair (\%)$} & \colhead{$f_M (\%)$} & \colhead{$\langle \mu_* \rangle$} & \colhead{$f_{\textrm{Q}} (\%)$} & \colhead{$f_{\textrm{IS}} (\%)$} \\
\colhead{} & \colhead{mass range} & \multicolumn{5}{l}{} & \colhead{$= f_M / \fpair$} & \colhead{SSFR / $(U-V)_{\textrm{cor}}$} & \colhead{  }}
\tablecomments{$N_p$, $N_r$, and $N_g$ are the number of observed projected pairs, the expected number of these that are chance alignments, and the number of primary galaxies, respectively. $\fpair = (N_p - N_r) / N_g$ is the number of physical secondaries per primary galaxy, of which a fraction $f_{\textrm{Q}}$ are quiescent (as determined using the two methods described in the text) and a fraction $f_{\textrm{IS}}$ are included in the primary sample (i.e., quiescent and massive). $f_M$ is the mean stellar mass in physical secondaries as a fraction of the host. Uncertainties in $\fpair$ reflect Poisson noise in $N_p$ and $N_r$; for other quantities, uncertainties are determined from bootstrap resampling of the primary galaxies.\label{tab:pairs}}
\startdata
\multicolumn{10}{l}{\emph{Quiescent galaxies}} \\
$0.4 < z < 1$ & $10.7 < \log M_*$ & 41 &9.7 & 177 & $18 \pm 4$ & $6.2 \pm 1.6$ & $0.35 \pm 0.06$ & $71 \pm 10 / 91 \pm 8$  & $ 35 \pm 12$ \\
$1.0 < z < 1.5$ & $10.7 < \log M_*$ & 27 &12.4 & 117 & $13 \pm 5$ & $5.3 \pm 1.7$ & $0.43 \pm 0.08$ & $66 \pm 16 / 78 \pm 18$  & $ 33 \pm 17$ \\
$1.5 < z < 2$ & $10.7 < \log M_*$ & 30 &11.8 & 100 & $18 \pm 6$ & $7.5 \pm 2.7$ & $0.41 \pm 0.06$ & $33 \pm 15 / 38 \pm 16$  & $ 18 \pm 14$ \\
$0.4 < z < 2$ & $10.7 < \log M_* < 10.9$ & 39 &18.0 & 141 & $15 \pm 5$ & $5.8 \pm 1.8$ & $0.39 \pm 0.07$ & $52 \pm 14 / 71 \pm 16$  & $\ldots$ \\
$0.4 < z < 2$ & $10.9 < \log M_* < 11.2$ & 48 &16.5 & 174 & $18 \pm 5$ & $6.5 \pm 1.8$ & $0.36 \pm 0.05$ & $47 \pm 12 / 67 \pm 13$  & $\ldots$ \\
$0.4 < z < 2$ & $11.2 < \log M_*$ & 22 &4.9 & 80 & $21 \pm 6$ & $8.6 \pm 2.9$ & $0.40 \pm 0.08$ & $76 \pm 13 / 82 \pm 12$  & $\ldots$ \\
\multicolumn{10}{l}{\emph{Star-forming galaxies}} \\
$0.4 < z < 1$ & $10.7 < \log M_*$ & 15 &4.3 & 84 & $13 \pm 5$ & $4.9 \pm 2.2$ & $0.39 \pm 0.07$ & $50 \pm 23 / 71 \pm 21$  & $\ldots$ \\
$1.0 < z < 1.5$ & $10.7 < \log M_*$ & 30 &14.5 & 127 & $12 \pm 5$ & $4.1 \pm 2.0$ & $0.34 \pm 0.08$ & $41 \pm 20 / 53 \pm 21$  & $\ldots$ \\
$1.5 < z < 2$ & $10.7 < \log M_*$ & 30 &12.9 & 108 & $16 \pm 6$ & $5.7 \pm 2.4$ & $0.36 \pm 0.08$ & $29 \pm 15 / 59 \pm 17$  & $\ldots$
\enddata
\end{deluxetable*}

From the point of view of galaxy assembly, an equally useful quantity is the amount of stellar mass contained in physical companions. We estimate this simply by computing the mean total stellar mass in projected secondaries, expressed as a fraction of the host mass, and subtracting the random contribution as described previously. We denote this quantity $f_M$. Averaged over all masses and redshifts, we find $f_M = 0.060 \pm 0.011$. The mean mass ratio $\langle \mu_* \rangle = f_M / \fpair$ is very nearly constant at $\approx 0.39$ in all redshift and primary mass ranges. As many authors have noted, this implies that the stellar mass delivered in mergers arrives primarily in more massive secondaries \citep[e.g.,][]{Hopkins10c}. In Section 5, we compare $\langle \mu_* \rangle$ to theoretical expectations.

Although we concentrate on the growth of quiescent galaxies for the remainder of this paper, for comparison with future work we also tabulate the corresponding quantities for star-forming galaxies in Table~\ref{tab:pairs}. 

\begin{figure*}
\centering
\includegraphics[width=0.45\linewidth]{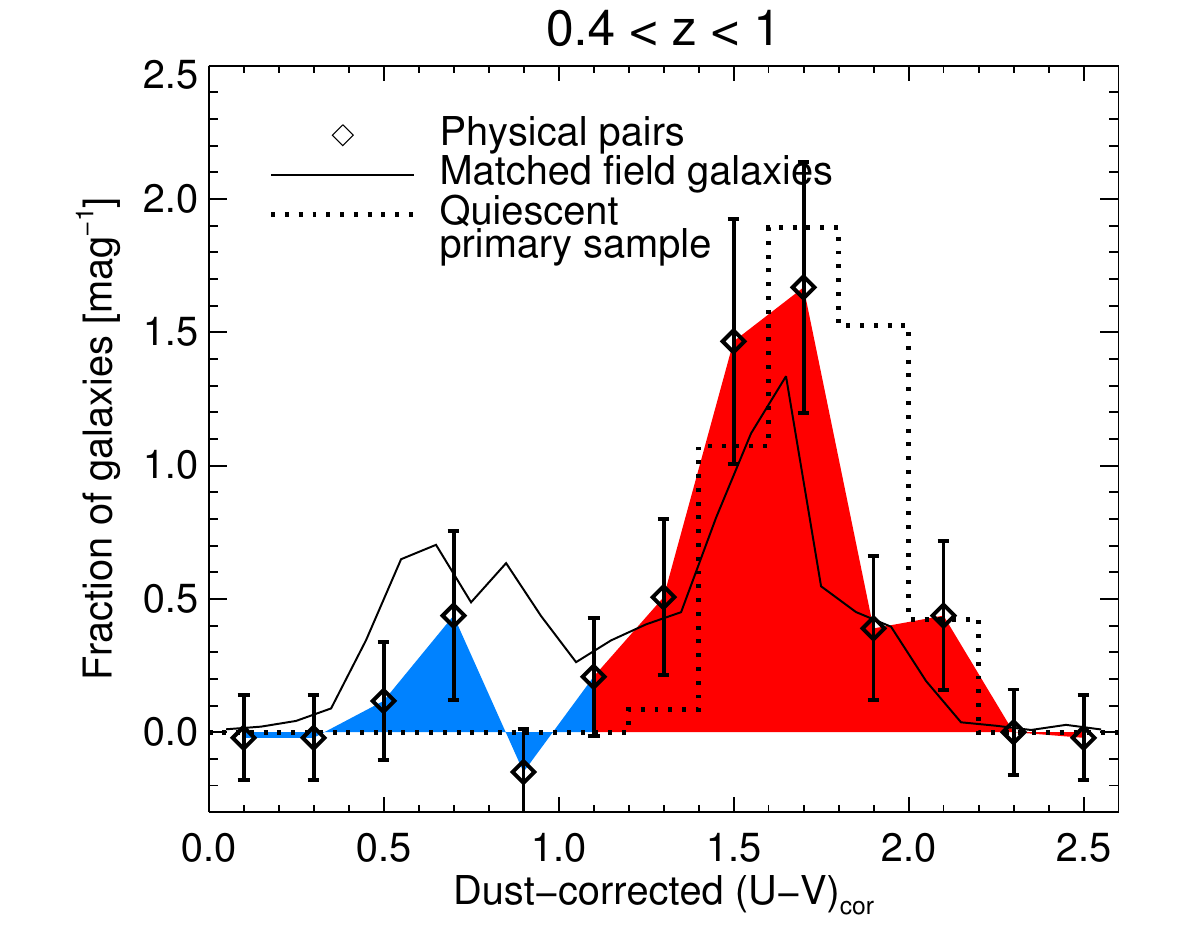}
\includegraphics[width=0.45\linewidth]{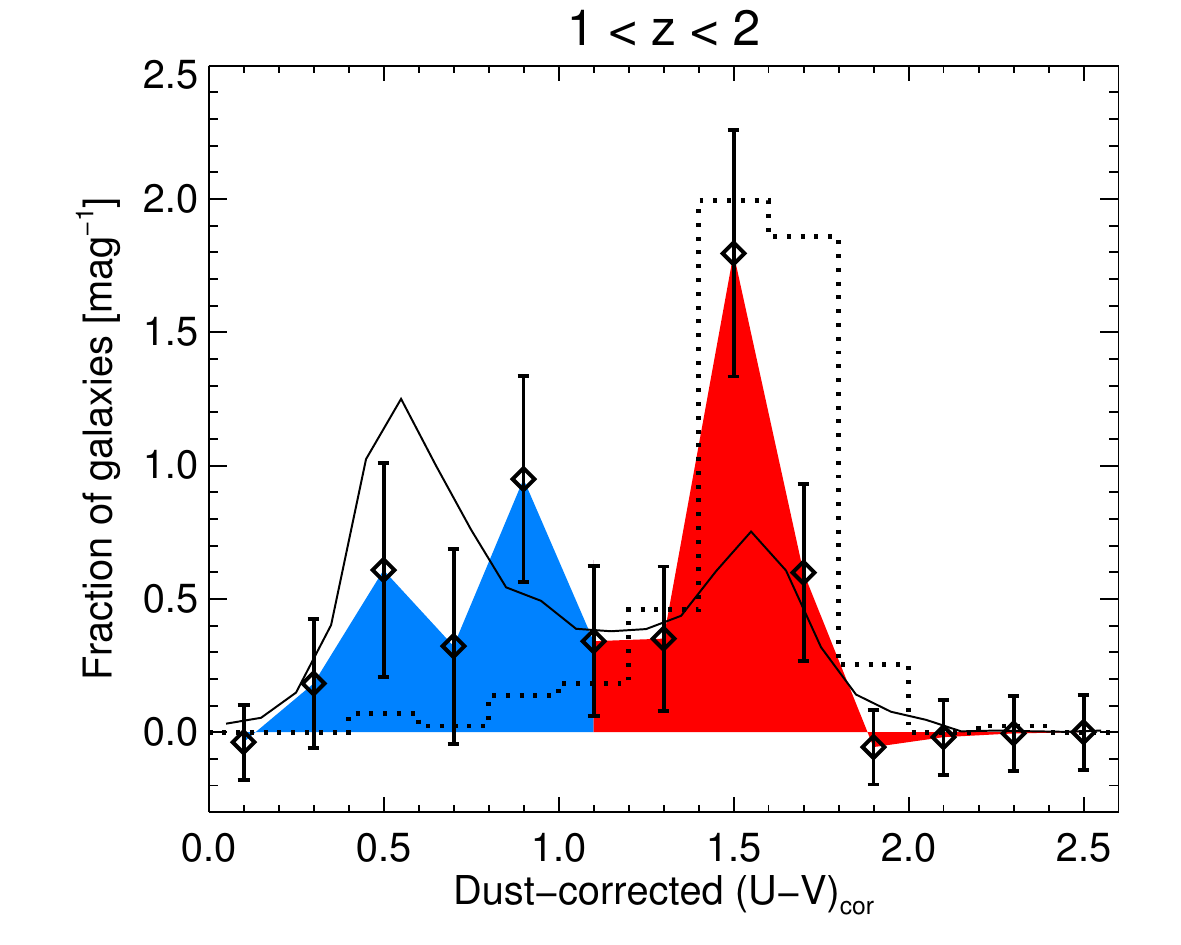}
\caption{Distribution of rest-frame colors of physical secondaries around the massive, quiescent galaxy sample in two redshift bins, compared to a field sample that is matched in stellar mass and redshift (solid) and to the primary sample (dotted). At higher redshifts, a significant fraction of companions are blue. \label{fig:colors}}
\end{figure*}

\subsection{Colors of Physical Secondaries}

Since physical secondaries likely represent the ``building blocks'' for the future mass assembly of quiescent galaxies, particularly in their outer regions, it is interesting to consider their stellar populations in relation to those of their hosts. In particular, the fraction of mergers which are ``dry'' (gas-poor) versus ``wet'' is an important input to models of galaxy evolution. Table~\ref{tab:pairs} presents the fraction $f_\textrm{Q}$ of physical secondaries which are quiescent. We calculate this using two definitions of quiescence: the $\SSFR < 0.02~\textrm{Gyr}^{-1}$ threshold used throughout this paper (also excluding MIPS detections), and a color selection $(U-V)_{\textrm{cor}} > 1.1$. Here
\begin{equation}
(U-V)_{\textrm{cor}} = (U-V)_{\textrm{rest}}-0.47A_V
\end{equation}
represents the \emph{extinction-corrected} rest-frame $U-V$ color \citep{Brammer09}. Overall, the two selections are qualitatively consistent: most physical companions to quiescent galaxies are themselves quiescent at $z < 1$, and this fraction decreases with redshift. 

This is illustrated in Figure~\ref{fig:colors}, which shows the color distribution of the physical secondaries. (As throughout, we have subtracted  the color distribution of similarly-selected galaxies in randomly placed apertures.) The secondaries are compared to a field sample with matched distributions in stellar mass and redshift (solid line) and to the primary quiescent host sample (dotted). In both redshift bins, the physical secondaries are on average redder than the field comparison sample. The fraction of blue secondaries increases with redshift, suggesting that the reservoir of future merger candidates includes progressively more gas-rich galaxies at earlier times. The implications for the merger descendants are interesting but not completely clear. On the one hand, secondary bursts of star formation are observed in spectroscopic samples of early-type galaxies at $z \simeq 1$ \citep{Treu05}. On the other hand, as we review in Section 5, merger timescales are expected to be $\gtrsim 1$~Gyr. If the processes driving satellite quenching are mostly confined to the final $\sim$Gyr, many of these blue secondaries may be much redder by the time of the final merger. Nevertheless, it seems likely that a significant fraction of mergers at $z \gtrsim 1$ are not completely dry, even for red hosts.

\begin{figure}
\centering
\includegraphics[width=0.9\linewidth]{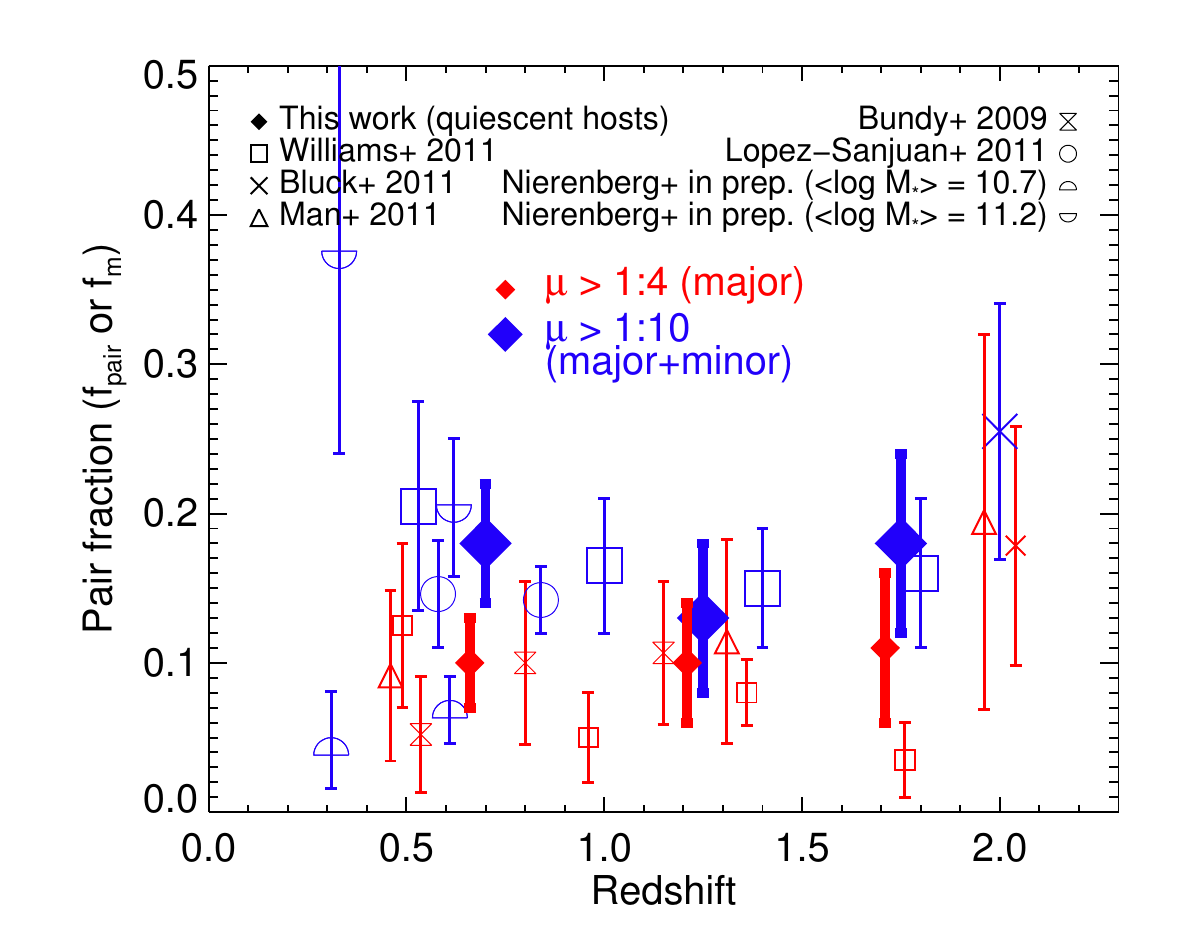}
\caption{Major ($\mu > 0.25$, small red symbols) and total ($\mu > 0.1$, large blue symbols) pair fractions are compared to recent independent measurements at $z < 2$. Here $\mu$ refers to either a flux or stellar mass ratio. Overall there is reasonable consistency within the statistical uncertainties, with no clear sign of strong evolution in $\fpair$ over $z = 0.5-2$. Expected systematic differences arising from different selection techniques are discussed in the text. Points are slightly offset in redshift for clarity.\label{fig:fpair_z}}
\end{figure}

\subsection{Comparison with Previous Work}

Comparisons to independent estimates of the pair fraction are complicated by the intrinsic differences in samples selected by various means (stellar mass, color, luminosity). In particular, as we discuss below, samples in which satellites are selected based on their stellar mass will systematically differ from those based on luminosity, particularly in the rest-frame optical. An advantage of the present study is the characterization of the merger rate and size growth using a uniform mass-based selection. Nevertheless, it is valuable to compare our $\fpair$ measurements to other works. In the following, we rescale published $\fpair$ measurements to our search area by assuming that $\fpair(R<R_{\textrm{max}}) \propto R_{\textrm{max}}$ \citep{Kitzbichler08,LopezSanjuan11}. Figure~\ref{fig:fpair_z} shows this comparison, focusing primarily on those studies that adopted a mass-based selection, included minor mergers, or probed to $z \simeq 2$. For comparison, we also plot the ``major'' pair fraction in our sample, defined by $0.25 < \mu_* < 1$.

At $z \lesssim 1$, our measurements are broadly in agreement with previous results when analogous samples are compared and search apertures are matched \citep[e.g.,][]{Kartaltepe07,Lin08,Rawat08,deRavel09}. Of particular interest is the comparison to \citet{LopezSanjuan11}, who identified minor spectroscopic pairs. Figure~\ref{fig:fpair_z} shows results for their red host sample with secondaries having rest-$B$ luminosity ratios $\mu_B > 0.1$. A.~Nierenberg et al.~(2012, in preparation, see also \citealt{Nierenberg11}) identify minor companions to spheroidal hosts split into two bins of stellar mass (Chabrier IMF), identifying satellites with flux ratios $\mu_{\textrm{F814W}} > 0.1$. The strong mass dependence they find highlights the importance of matching hosts in stellar mass when comparing pair fractions or analyzing the associated size growth. We note that Nierenberg et al.~use a local background estimation that is expected to yield smaller raw pair fractions. \citet{Bundy09} selected major pairs at $z < 1.4$ having $K$-band flux ratios $\mu_K > 0.25$. Their results for red hosts with $\log M_* > 10.5$ (Chabrier IMF) are shown in Figure~\ref{fig:fpair_z}. Several authors have inferred merger rates from morphological signatures (e.g., \citealt{Lotz08b}, \citealt{Conselice09}, \citealt{Bridge10}). For a recent review, we refer to \citet{Lotz11}.

Few other studies have considered minor mergers at $z \gtrsim 1$. Among these, \citet{Williams11} is the most directly comparable to our work, as their selection is based on stellar mass. Figure~\ref{fig:fpair_z} shows their quiescent, $\log M_* > 10.8$ (Kroupa IMF) host galaxy sample, for which they find $\fpair \approx 0.16-0.20$ ($\mu_* > 0.1$) essentially independent of redshift, in encouraging agreement with our results.

\citet{Man11} use $H$-band \emph{HST}/NICMOS imaging to assess the major pair fraction for massive ($\gtrsim 10^{11} \msol$) hosts, selecting secondaries with $H$-band flux ratios $\mu_H > 0.25$. Recently, \citet{Bluck11} (see also \citealt{Bluck09}) studied fainter companions around a similarly massive population by identifying close pairs to a limiting flux ratio of $\mu_H = 0.01$ in NICMOS imaging. As Figure \ref{fig:fpair_z} shows, our data are consistent with these flux-based selections at $z \lesssim 1.7$. There is a hint that the major and total ($\mu > 0.1$) pair fractions rise toward $z \simeq 2$, but this is not very significant at present. (\citealt{Bluck11} demonstrate stronger increases toward $z \simeq 3$.) Furthermore, samples in which secondaries are selected based on rest-optical flux will not agree in detail with stellar mass-based samples (see \citealt{Bundy04}). At $z \simeq 2$, the $H$ band probes the rest-frame $V$ band. A significant dispersion in the stellar mass-to-light ratio $M_*/L$ is thus expected, and the mean $M_*/L$ declines substantially with decreasing mass. Assuming a constant $M_*/L$ equal to that of the host and a limiting flux ratio $\mu_H > 0.1$, for example, will include bluer galaxies with lower mass ratios $\mu_* < 0.1$, likely resulting in an elevated $\fpair$ compared to a mass-selected sample. This effect is expected to become stronger toward lower mass ratios and toward higher redshifts as the $H$ band probes bluer rest wavelengths, possibly impacting trends with redshift in flux-selected samples.

Finally, \citet{vanDokkum10} used a novel method to infer indirectly the rate at which massive galaxies assemble mass through mergers. They tracked the stellar mass growth of a sample over $z=0-2$ with constant comoving number density and subtracted an estimate of the in situ star formation. Their estimated ``specific assembly rate'' is $\dot{M_*}/M_* = 0.03 (1+z)$~Gyr${}^{-1}$. At $z \sim 1$ this compares well with our pair counting estimate of $f_M / \tau_e \approx 0.07 / \tau_e$ for merger timescales $\tau_e \sim 1$~Gyr (see Section 5.2), although we find a weaker redshift dependence.

\section{Connecting Size Growth with Mergers}

In this section we present simple models that compare the rate of size growth measured in Section~3 with that attributable to mergers of close pairs as studied in Section~4. Sections 5.1 and 5.2 review the theoretical ingredients necessary to convert pair fractions into growth rates in mass and size. In Section 5.3 we examine the growth in the mean size of quiescent galaxies and discuss how the rate of size evolution experienced by any individual galaxy may be substantially smaller. Finally, in Section 5.4 we combine our constraints on the distribution of sizes of quiescent galaxies with the evolution of their number density to establish a minimum growth rate, which we then compare to a merger model.

\subsection{Merger Timescales and the Distribution of Mass Ratios\label{sec:timescales}}

Converting the observed number of physical companions into a merger rate requires us to specify the timescale during which a merger appears within our search aperture, i.e., a projected separation between 10 and 30 $h^{-1}$~kpc. We define an effective timescale $\tau_e$ that incorporates two physical effects. The first is the mean time $T_{\textrm{mg}}$ during which a bound, sinking satellite appears within our search aperture. As discussed in Section 4, however, not all of the physical secondaries we counted are necessarily bound to their host. By subtracting the number of pairs found in randomly placed apertures, we account for interlopers in the far foreground and background of the galaxies in our primary sample, but we can expect the remaining ``physical'' secondaries to include both bound satellites \emph{and} other galaxies in the larger group-scale environment that are not bound. As is common practice in merger rate studies, we account for this by defining a factor $C_{\textrm{mg}}$ (see also \citealt{Bundy09}) to represent the fraction of physical secondaries that are bound and due to merge on a typical timescale $T_{\textrm{mg}}$. The effective timescale is then $\tau_e = T_{\textrm{mg}} / C_{\textrm{mg}}$.

\citet{Patton08} study projected pairs of similar luminosity in the SDSS. They assume that mergers of luminous pairs occur with a typical timescale of $T_{\textrm{mg}} = 0.5$~Gyr. Based on tests using the Millennium simulation \citep{Springel05}, they estimate $C_{\textrm{mg}} \approx 0.5$ (their $f_{\textrm{3D}}$) for the most luminous pairs, resulting in an effective timescale $\tau_e = T_{\textrm{mg}} / C_{\textrm{mg}} \approx 1.0$~Gyr.

\citet{Lotz08,Lotz10a,Lotz10b} investigate merger timescales for disk galaxies using high-resolution hydrodynamical simulations. For self-similar mergers of their most massive disk G3, they find a mean timescale $\langle T_{\textrm{mg}}\rangle = 0.7$~Gyr within our adopted search annulus. Similar to \citet{Patton08}, \citet{Lotz11} allow for projection effects by setting $C_{\textrm{mg}} = 0.6$, resulting in an effective $\tau_e = 1.2$~Gyr.

\citet{Kitzbichler08} calibrated $\tau_e$ using the Millennium simulation, coupled with a semi-analytic model (SAM) of galaxy merging and evolution \citep{deLucia07}. Considering major mergers ($0.25 < \mu_* < 4$) of $M_*=10^{11} \msol$ galaxies and a search aperture of $R < 30~h^{-1}$~kpc, they find $\tau_e = 2.7$~Gyr. Since $\tau_e$ scales approximately as the outer radius of the aperture, and we exclude the inner $10~h^{-1}$~kpc, the appropriate timescale for our study would be $\sim 2/3$ of this, or $\tau_e \approx 2~\textrm{Gyr}$. 

\begin{figure}
\centering
\includegraphics[width=0.9\linewidth]{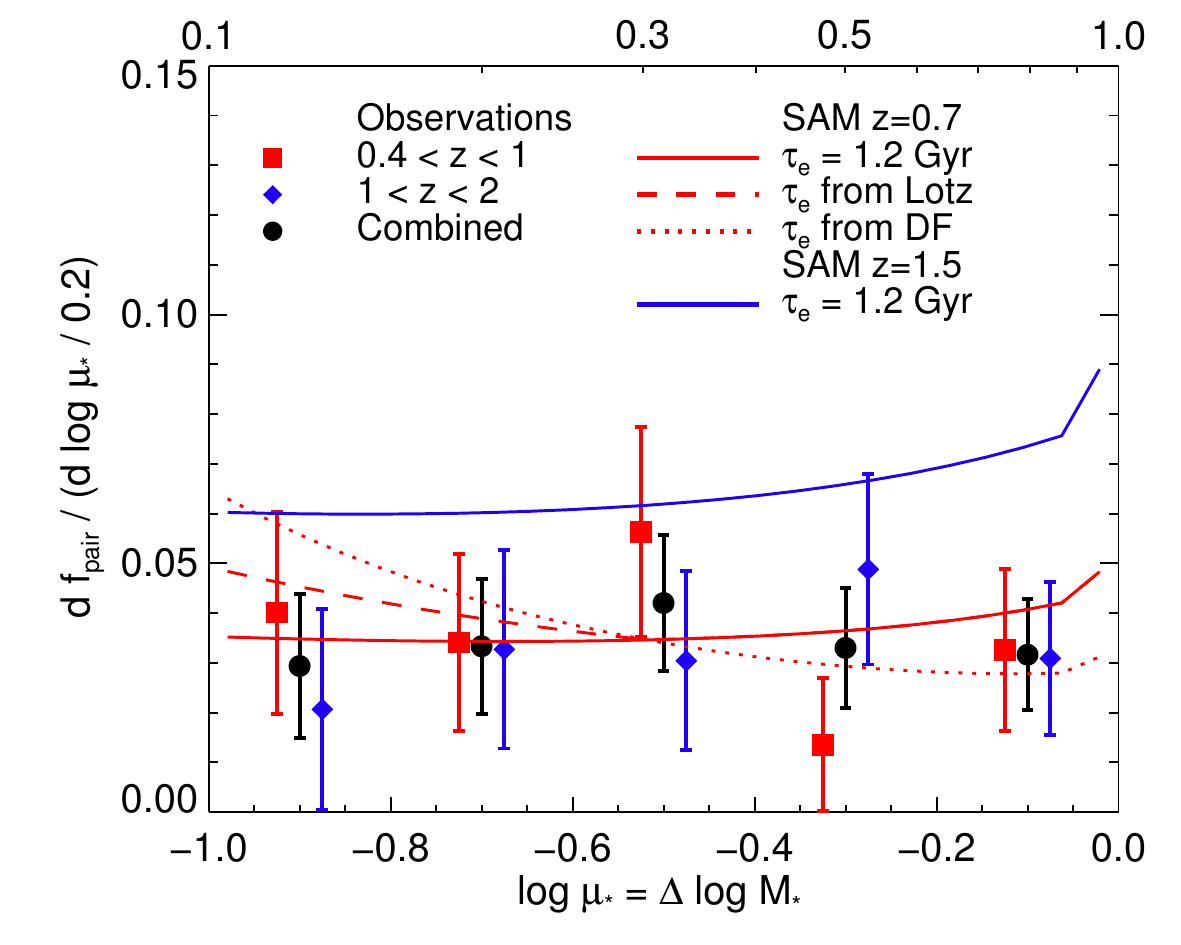}
\caption{Distribution of the stellar mass ratio $\mu_*$ of physical secondaries around $\log M_* > 10.7$ quiescent galaxies in two redshift bins. Merger rate predictions of the \citet{Hopkins10c} SAM are overlaid under several assumptions for the timescale $\tau_e$: a constant $\tau_e = 1.2$~Gyr (solid line), $\tau_e \propto \mu_*^{-0.3}$ for $\mu_* < 0.3$, in line with the Lotz simulations (dashed), and $\tau_e \propto [\mu \log (1+\mu^{-1})]^{-1}$ (dotted) as expected from pure dynamical friction considerations.\label{fig:pairfrac}}
\end{figure}

Taken together, these studies imply $\tau_e = 1-2$~Gyr for major mergers. In order to make progress in the present study, we also need an estimate of $\tau_e$ for minor mergers, which will likely be larger. The outcome will depend on the relative contribution of minor mergers, i.e., the distribution of $\mu_*$ in our sample, and on how strongly the timescales vary with $\mu_*$. Figure~\ref{fig:pairfrac} breaks the observed pair fractions from Section 4 into several bins of the mass ratio $\mu_*$. The distribution is essentially flat in $\log \mu_*$; this corresponds to a mass function that rises as $\mu_*^{-1}$. We note that a uniform distribution over $-1 < \log \mu_* < 0$ has a mean $\langle \mu_* \rangle = 0.39$, in agreement with our measurement of $\langle \mu_* \rangle = 0.39 \pm 0.04$ (Section 4).

To understand the flat distribution in Figure~\ref{fig:pairfrac}, we compare the measured pair fractions to merger rates $R_{\textrm{mg}} = \fpair / \tau_e$ predicted by the SAM of \citet{Hopkins10c} under three choices of the observability timescale $\tau_e$ described in the caption. Within the uncertainties, our observations are consistent with any of these timescale scalings.  At higher redshifts, the SAM predicts higher $\fpair$ (blue curve), which we do not observe. However, the distribution in $\mu_*$ maintains the same shape nearly independently of mass or redshift. This is not particular to the \citet{Hopkins10c} model, but is a generic feature of many SAMs \citep[see][]{Lotz11}. The SAM also provides an estimate of the amount of stellar mass in $\mu_* < 0.1$ mergers that we do not probe observationally. Only $\simeq 9\%$ of the predicted mass assembly rate (i.e., the rate at which stellar mass is delivered through mergers) is due to $\mu_* < 0.1$ mergers. This simply reflects the fact that for such low mass ratios, the time for a galaxy to descend from the virial radius to the center quickly exceeds a Hubble time \citep{Taffoni03,BoylanKolchin08}. By observing $\mu_* > 0.1$ pairs, therefore, we expect to account for the vast majority of the mass assembly.

In summary, the effective timescale for major mergers is likely $\tau_e = 1-2$~Gyr. For lower mass ratios, estimates are even less certain. However, since the physical secondaries are not overly dominated by the lowest mass ratio systems, consistent with the predictions of SAMs, we expect the appropriate average $\tau_e$ for our sample to be only moderately higher. In the following analysis we present results for models spanning a range of timescales.

\subsection{Size Growth Efficiency\label{sec:alpha}}

In order to address whether mergers drive the observed size growth, we need to know how the half-light radius $R_h$ of a galaxy changes after undergoing a merger of mass ratio $\mu$. This question has been addressed in the literature both analytically and using extensive suites of merger simulations. The growth efficiency is commonly parameterized by $\alpha = d \log R_h / d \log M_*$. For $1:1$ mergers of spheroids, both the mass and radius approximately double and $\alpha \simeq 1$ \citep[e.g.,][]{Hernquist93,Nipoti03,BoylanKolchin06}.

Simple virial arguments based on energy conservation show that the growth efficiency can be higher for more minor mergers \citep{Hopkins09,Bezanson09,Naab09}. Assuming that the orbit is parabolic, that the progenitors and merger product are structurally homologous, and that there is negligible energy transfer from the stars to the dark halo,
\begin{equation}
\alpha = 2 - \frac{\log(1+\mu^{2-\beta})}{\log(1+\mu)},\label{eqn:alpha}
\end{equation}
where we have assumed the progenitors lie on a $R_h \propto M_*^{\beta}$ relation. For self-similar mergers $\mu = 1$ and we recover $\alpha = 1$. For a mass--radius slope of $\beta = 0.57$ (Section 3) and the lowest mass ratios we observe ($\mu = 0.1$), this estimate becomes $\alpha =  1.6$. We therefore expect an appropriately averaged $\langle \alpha \rangle$ over the mass ratios we consider to lie in this range.

Recognizing the assumptions entering this simple formula, it is essential to verify its predictions with merger simulations. \citet{Nipoti09L} simulated hierarchies of multiple dry minor mergers of spheroids and found $\langle \alpha \rangle =1.30$. More recently, \citet{Nipoti11conf} performed a suite of $\mu=0.2$ dry spheroid mergers and found $\langle \alpha \rangle = 1.60$ (see also C.~Nipoti et al 2011, submitted). \citet{Oser11} investigated the relevance of Equation~\ref{eqn:alpha} in a cosmological hydrodynamical simulation and found it to be accurate. Altogether, based on these results, we consider $\alpha \sim 1.3 - 1.6$ to be a reasonable average over the mass ratios we consider. We note that a higher efficiency  ($\alpha > 1.6$) has not been demonstrated, when averaged over a representative set of orbits, in any $N$-body simulation of which we are aware.

\subsection{Matching the Observed Growth of the Quiescent Population\label{sec:simple}}

\begin{figure*}
\centering
\includegraphics[width=0.45\linewidth]{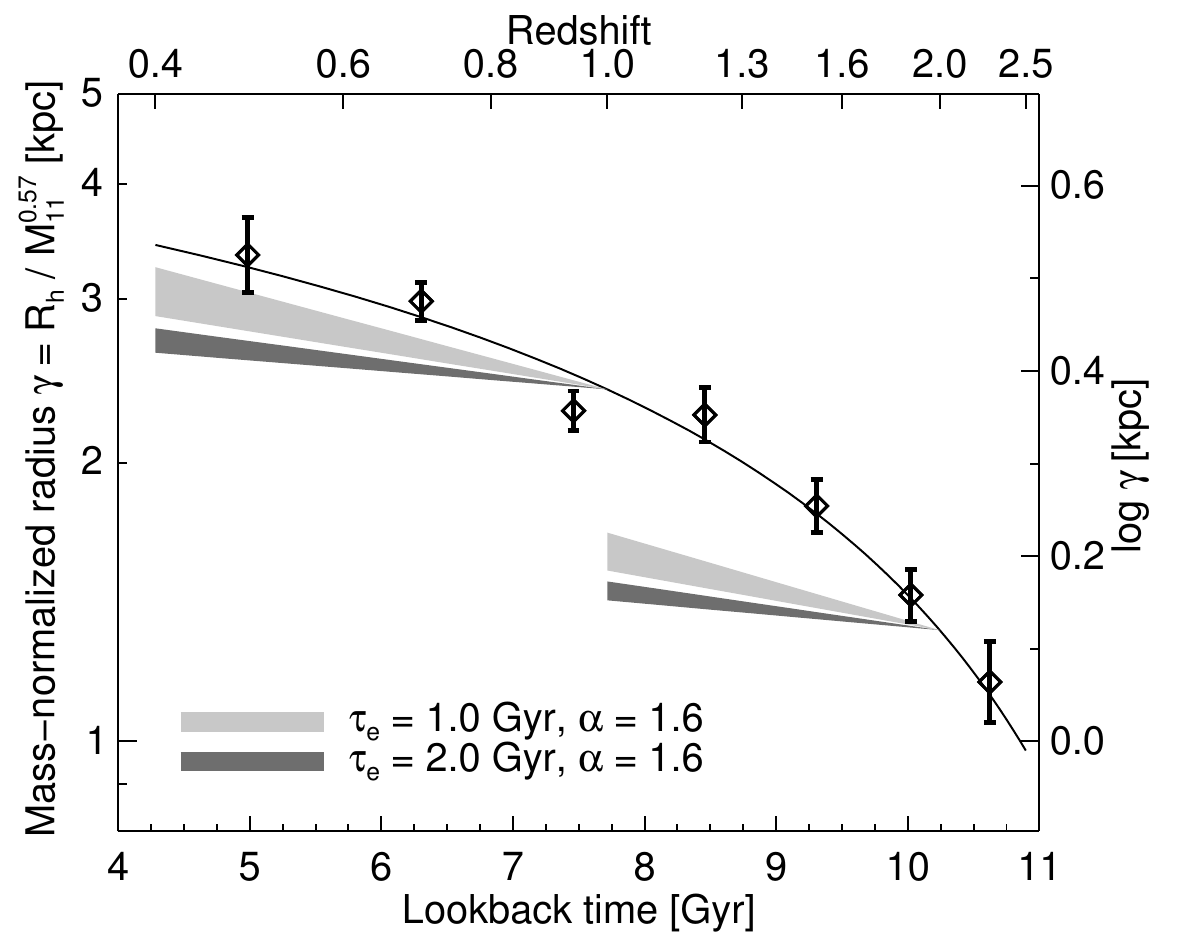}
\includegraphics[width=0.45\linewidth]{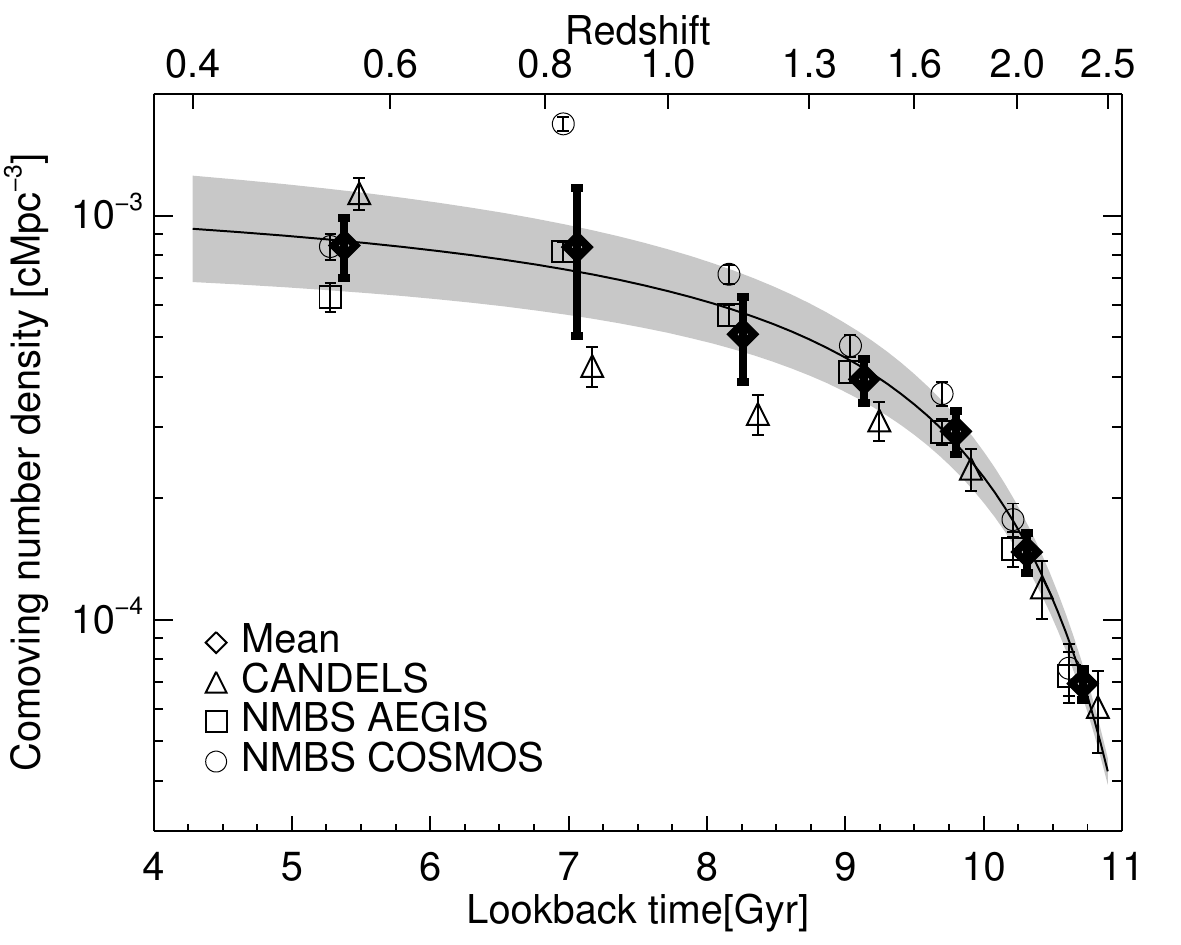}
\caption{\textbf{Left:} Observed evolution in size measured at fixed mass for quiescent galaxies, reproduced from Figure~\ref{fig:oned_growth}, compared to predicted trajectories expected from the simple merging model discussed in the text. We begin with progenitors at $z = 2$ and 1 and predict their future growth. The width of the shaded bands corresponds to the uncertainty in $f_M$. At $z \lesssim 1$, mergers plausibly account for much of the size growth if the timescale is short, but at higher redshifts they cannot match the rapid growth of the population. \textbf{Right:} Comoving number density of $\log M_* > 10.7$ quiescent galaxies in CANDELS and the NMBS. Error bars for individual fields (slightly offset in redshift for clarity) reflect Poisson noise only, while the error in the mean (diamonds) is determined empirically from the scatter among the fields. The solid line and shaded region indicate the fit used in our model and its $1\sigma$ uncertainty.\label{fig:simple_growth}}
\end{figure*}

With estimates for the growth efficiency $\alpha$ and merger timescale $\tau_e$ in hand, we can now proceed to a simple model that estimates the rates of growth in mass and size due to mergers. In a time interval $\Delta t$, the stellar mass of the average quiescent host in our sample increases by $\Delta \log M_* = \log (1+f_M)^{\Delta t / \tau_e}$, while the radius by definition increases by $\alpha \Delta \log M_*$. Since we expect $\alpha > \beta = 0.57$, as discussed in Section 5.1, mergers will shift the mean mass-size relation:
\begin{align}
\Delta \log \gamma &= \Delta \log R_h - \beta \Delta \log M_* = (\alpha - 0.57) \Delta \log M_* \nonumber \\
 & = (\alpha - 0.57) \log (1 + f_M)^{\Delta t/\tau_e} \label{eqn:growth}
\end{align}
We have neglected here the small change in the number density arising from mergers within the sample over the interval $\Delta t$. This incurs a fractional error of $\sim f_{\textrm{IS}} \fpair \approx 5\%$ in the mass accreted, negligible for our purposes.

Figure~\ref{fig:simple_growth}a reproduces the observed growth in $\gamma$ from Figure~\ref{fig:oned_growth}. Using Equation \ref{eqn:growth}, we overlay growth trajectories for representative values of the growth efficiency and merger timescale to illustrate the evolution of the quiescent galaxy populations in place at $z = 2$ and $z = 1$. Here we have taken $f_M$ appropriate to $z = 0.4 - 1$ and $z = 1-2$ (Table~\ref{tab:pairs}) and applied a 15\% correction to account for both additional satellites below our $\mu_* = 0.1$ limit and for possible catastrophic redshift errors (Section~\ref{sec:photoz}). We assume that all galaxies grow smoothly at this rate, i.e., we do not incorporate stochasticity in the incidence of mergers.

The primary conclusions from Figure~\ref{fig:simple_growth}a are twofold. First, at $z \lesssim 1$ the pairs we observe can plausibly account for most of the observed size growth if an effective timescale $\tau_e \sim 1$~Gyr, at the short end of the estimates discussed in Section~\ref{sec:timescales}, and an average growth efficiency $\alpha \approx 1.6$ are valid. Second, at $z \gtrsim 1$ the observed growth in $\log \gamma$ per unit time increases significantly. This enhanced growth rate cannot be matched by mergers using any reasonable choices of $\tau_e$ and $\alpha$.

As discussed in Section 3, however, an important objection to the model comparisons in Figure~\ref{fig:simple_growth}a is that we are tracking the mean growth rate of the entire population, as if all sources are enlarged in lockstep. In reality, the population at any redshift comprises both old galaxies which formed at higher redshift and which presumably are growing via mergers, {\it and} sources newly arriving on the quiescent sequence, whose size may reflect their epoch of formation. Galaxies appearing later are typically formed from gas-poorer progenitors. They are therefore expected theoretically to experience less dissipation in their formation, possibly leading to less compact remnants \citep{Robertson06,Khochfar06,Hopkins10b,Shankar11}. 

Figure~\ref{fig:simple_growth}b demonstrates that the \emph{comoving number density} of quiescent galaxies increases very rapidly at $z \gtrsim 1.3$, exactly where the growth in mean size is most rapid. For example, only $\sim 25\%$ of the sample at $z \sim 1$ was already formed and quiescent at $z \sim 2$. These early galaxies may need only to grow marginally into the compact tail of the distribution at $z \sim 1$. They might then experience significantly less growth than the population mean tracked in Figure~\ref{fig:simple_growth}a. In this figure, we have combined our CANDELS catalog with those from the NEWFIRM Medium Band Survey (NMBS; \citealt{Whitaker11}) to increase the total volume. Densities in the various fields agree closely at $z \gtrsim 1.5$, where large volumes are probed, while cosmic variance dominates at $z \lesssim 1.5$. 

There is observational support at $z \sim 0$ for the idea that younger early-type galaxies are larger at fixed mass \citep{Shankar09,vanderWel09,Bernardi10}. On the other hand, some recent studies at higher redshift have found no sign of such a correlation (\citealt{Trujillo11,Whitaker11b}; but see \citealt{Saracco11}). Although the true situation remains unclear, it is interesting to consider size growth assuming that the oldest galaxies at a given redshift and stellar mass are the smallest, since this corresponds to the minimum rate of growth that individual old galaxies must undergo. We now seek to construct a test that accounts for the continual emergence of quiescent systems.

\subsection{A Minimum Rate of Growth for Early Compact Galaxies}

The physical processes that determine the size of a galaxy in its early history might therefore be quite different from those that drive its subsequent growth. \citet{Oser10} described a ``two phase'' picture that, while obviously a simplification at some level, still provides a useful paradigm for galaxy growth. The first phase is characterized mainly by in situ star formation, while in the second phase, most growth occurs through accretion of stars. We wish to test whether mergers are sufficient to power size growth in this second phase. As we discussed in Section 5.3, growth in the mean size of the quiescent population (Figure~\ref{fig:simple_growth}a) entails processes operating in both phases which are hard to uniquely disentangle. Evolution in the population mean alone does not necessarily imply that any individual galaxy must grow in size. The key evidence for growth in the ``second phase'' is the declining abundance of compact systems. Observationally, we seek to explain the minimum rate at which high-$z$ compact galaxies must evolve so as to avoid leaving too many compact remnants at later times.

\begin{figure*}
\centering
\includegraphics[width=0.45\linewidth]{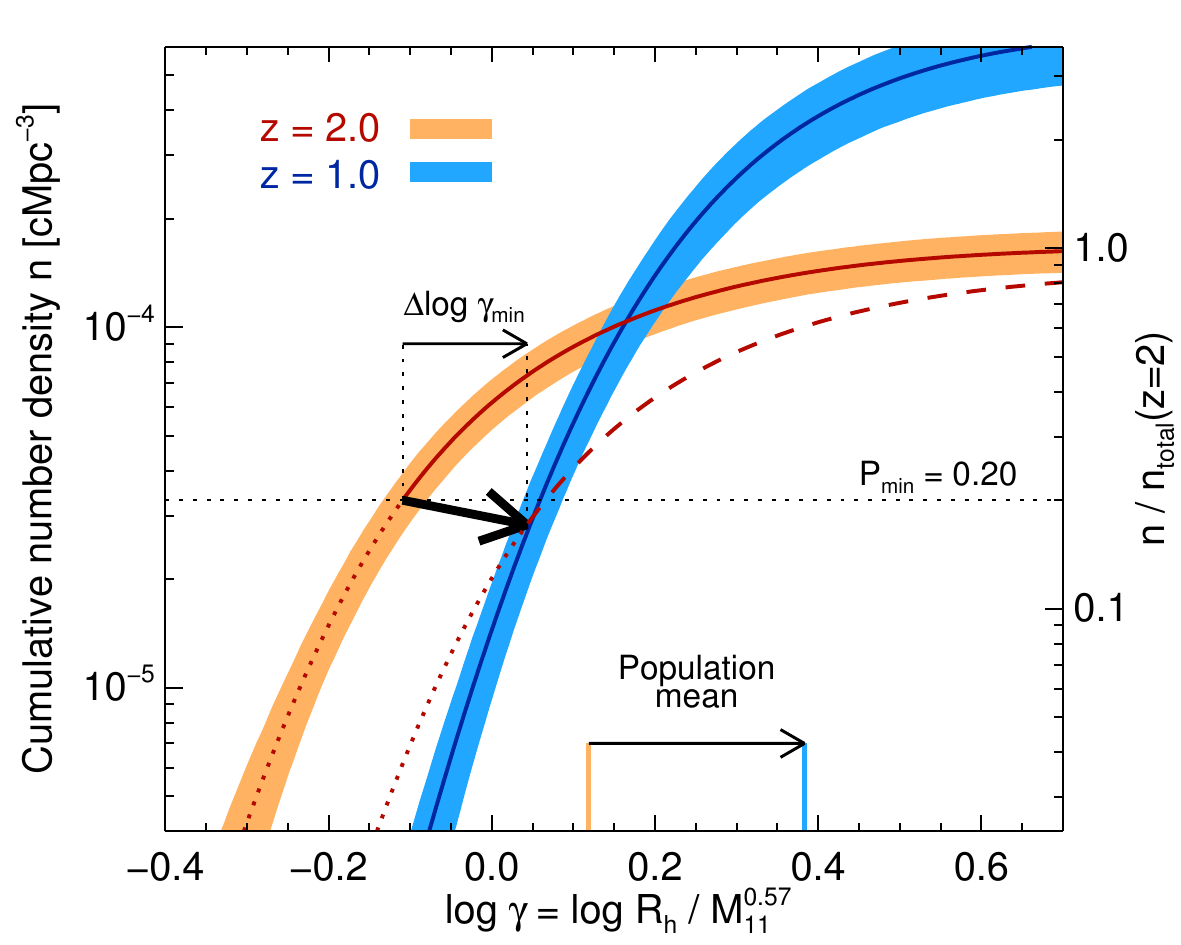}
\includegraphics[width=0.45\linewidth]{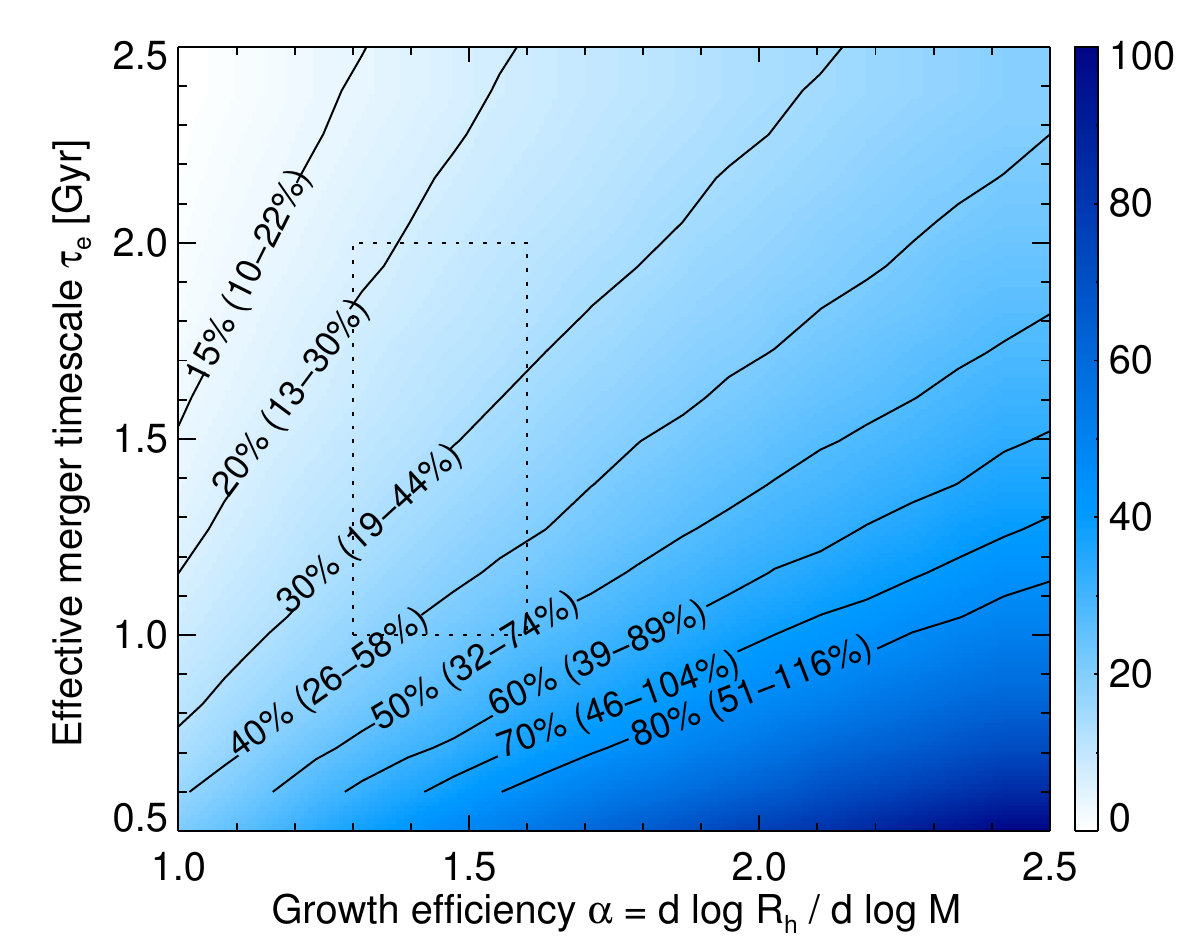}
\caption{\textbf{Left:} Cumulative compactness functions at $z=2$ and $z=1$, representing the comoving number density of quiescent galaxies more compact than a given $\gamma$ based on fits described in Section 3. The thick arrow indicates the minimum size expansion necessary to ensure that descendants of the $z = 2$ population can be accommodated within the $z=1$ distribution. As discussed in the text, we do not use the lowest 20\% of the $z = 2$ CF as a constraint. Bands indicate $1\sigma$ confidence regions. \textbf{Right:} The fraction $\Delta \log \gamma_{\textrm{merg}} / \Delta \log \gamma_{\textrm{min}}$ of the minimum required growth rate over $z = 1-2$ that is producible by mergers under various assumptions for the timescale and growth efficiency, with the $10-90\%$ confidence intervals in parentheses. The dotted box outlines the range of likely parameters discussed in Sections 5.1 and 5.2.\label{fig:demo}}
\end{figure*}

To test for growth in this second phase requires the \emph{distribution of sizes} at two redshifts and the \emph{relative abundances} of the progenitors and candidate descendants. We focus on the redshift interval $z=1-2$ to illustrate the method. Figure~\ref{fig:demo} shows the cumulative distributions of the mass-normalized radius $\gamma$ at $z = 2$ and $z=1$ using the fits presented in Section 3. These have been scaled to total number densities using the fit in Figure~\ref{fig:simple_growth}b. We term these compactness functions (CFs) in analogy to the more familiar stellar mass function (see Bezanson et al.~2011 for a demonstration in terms of inferred velocity dispersion).

Mergers will shift the $z = 2$ CF in two ways in Figure~\ref{fig:demo}. First, galaxies will expand according to Equation~\ref{eqn:growth}, which will shift the distribution rightward toward larger $\gamma$. We again assume that this growth is uniform and neglect stochasticity in mergers. A second, less important, effect is that some of these mergers will be among galaxies within the sample. This will reduce the number density of the sample over time, moving the size distribution parallel to the $\log n$ axis. In Section 4, we called these ``intrasample mergers'' and measured the fraction $f_{\textrm{IS}}$ of physical secondaries they represent. The rate of intrasample mergers is then $f_{\textrm{IS}}\fpair / \tau_e$. Bearing in mind that $f_{\textrm{IS}}\fpair \approx 0.03 - 0.06$ is small (Table \ref{tab:pairs}), we can approximate the resulting reduction in number density over an interval $\Delta t$ by
\begin{equation}
\Delta \log n \approx \log (1 - f_{\textrm{IS}} f_{\textrm{pair}})^{\Delta t/\tau_e}.
\end{equation}
We note that the secondaries removed from the sample in intrasample mergers will preferentially be low mass, but since $\gamma$ is defined to be statistically independent of $M_*$, it is a good approximation to shift the CF uniformly.

The evolution of the $z = 2$ CF will thus proceed rightward and slightly downward in Figure~\ref{fig:demo}, as indicated by the thick arrow 
that connects the $z = 2$ CF (solid red line) to that of its descendants (dashed).
The length of the arrow indicates the magnitude of the size evolution and depends on $\fpair$ and $\tau_e$. A plausible evolutionary path must shift the $z=2$ CF to lie below that observed at $z=1$, otherwise too many compact descendants would remain at $z = 1$. We use this to define a minimum growth rate for $z=2$ compact galaxies consistent with the observed depletion in the number density of similarly compact systems.

Before embarking on this task, it is necessary to define a minimum percentile $P_{\textrm{min}}$ of the $z=2$ CF that we wish to fit within the observed $z = 1$ distribution. For example, if we require only that the largest 30\% ($P_{\textrm{min}} = 0.7$) of the $z=2$ descendants fit within the $z=1$ distribution, then no size growth is necessary, as Figure~\ref{fig:demo} shows. At the other extreme, if we require that \emph{all} $z=2$ descendants are accommodated ($P_{\textrm{min}} = 0$), the minimum necessary growth is approximately the same as the difference in the population means at the two redshifts, which we considered in Section 5.3.

In practice, some intermediate $P_{\textrm{min}}$ must be chosen. Although smaller values of $P_{\textrm{min}}$ provide stronger constraints, this must be balanced against our wish not to extrapolate fits of the observed $\gamma$ distribution down to arbitrarily small $\gamma$, where they are poorly constrained by the finite number of galaxies. In the following we conservatively set $P_{\textrm{min}} = 0.2$, which is large enough that the empirical CFs are fairly well constrained at $z < 2$ (see Figure~\ref{fig:sizemodel}). The thick arrow in Figure~\ref{fig:demo} has the minimum length necessary to shift the 20th percentile of the $z = 2$ CF beneath the $z=1$ distribution. The corresponding growth $\Delta \log \gamma_{\textrm{min}}$ can be taken as the minimum amount of growth necessary to sufficiently deplete the abundance of compact systems. Following the discussion in Section \ref{sec:simple}, this minimum growth is less than the difference in the means of the two CFs indicated at the bottom of the panel. Errors on $\Delta \log \gamma_{\textrm{min}}$ are estimated by repeating this calculation using many samples from the Markov chains used to fit the $\gamma$ distribution (Section 3). Number densities are also randomly perturbed from the mean fit as illustrated by the gray band in Figure~\ref{fig:simple_growth}b.

Figure~\ref{fig:demo}a demonstrates that the minimum  growth over $z = 1-2$ is $\Delta \log \gamma_{\textrm{min}} = 0.16 \pm 0.03$, assuming a size growth efficiency of $\alpha = 1.6$. Throughout we take $\langle \mu_* \rangle = 0.39$ and also set $f_{\textrm{IS}} = 0.18$ appropriate to $z\sim2$ (see Table~\ref{tab:pairs}), although the results are extremely insensitive to  this value. This minimum growth can now be compared to that expected from mergers via Equation \ref{eqn:growth}: $\Delta \log \gamma_{\textrm{merg}} = 0.08 \pm 0.02$, assuming a short timescale of $\tau_e = 1$~Gyr. Therefore, it appears that only $\Delta \log \gamma_{\textrm{merg}} / \Delta \log \gamma_{\textrm{min}} \approx 50\%$ of the required growth over this interval can be attributed to mergers. For longer merger timescales or lower growth efficiencies, this fraction would be less.  In Figure~\ref{fig:demo}b we show how the fraction depends on $\tau_e$ and $\alpha$. The dotted line outlines the region of likely parameters discussed in Sections 5.1 and 5.2. We conclude that mergers alone are unlikely to achieve the minimum rate of expansion required between $z=2$ and $z=1$, even for favorable assumptions regarding these theoretical parameters.
\begin{figure}
\centering
\includegraphics[width=\linewidth]{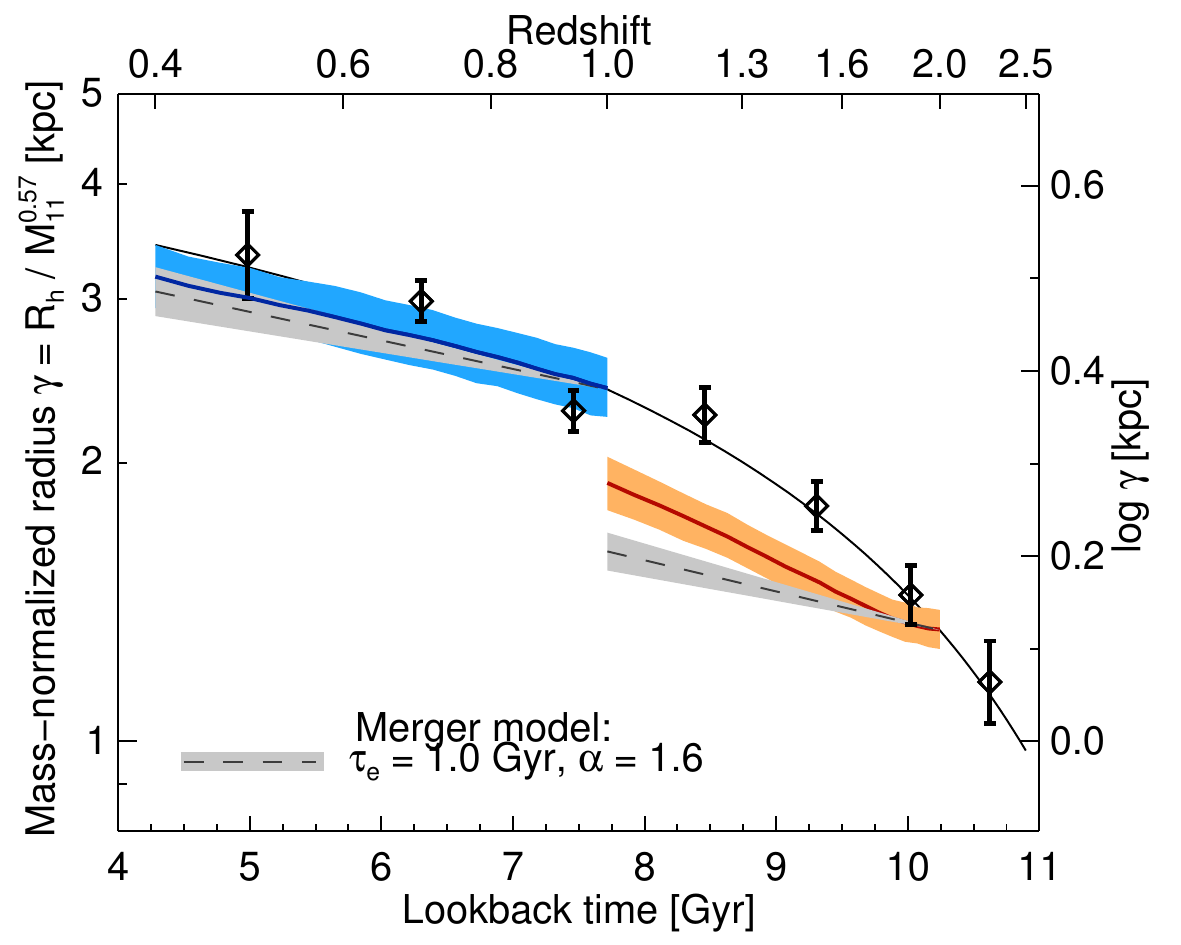}
\caption{Minimum required rates of growth for quiescent populations in place at $z=2$ and $z=1$, indicated by the dark gray bands (colored bands in online version), are compared to the growth rates of a simple merger model (light gray) discussed in the text. Black diamonds reproduce mean sizes from Figure~\ref{fig:simple_growth}a. The thickness of the merger trajectories reflects the uncertainty in $f_M$. All indicated uncertainties are $1\sigma$.  \label{fig:mingrowth}}
\end{figure}

The exercise can readily be repeated over other redshift intervals. Figure~\ref{fig:mingrowth} shows the minimum growth rate that progenitors at $z = 2$ and $z = 1$ must undergo to avoid leaving too many late compact remnants. This minimum rate is compared to the mean evolution of the quiescent population, introduced in Figure~\ref{fig:simple_growth}, and to the merger model predictions. During the period $z=1-2$ over which the number density is rapidly increasing, the population mean (black symbols) evolves more quickly than the minimum rate (red line). Both, however, exceed the expected growth rate from mergers (gray band), even for a favorable choice of $\tau_e = 1$~Gyr and $\alpha = 1.6$. From $z = 0.4-1$ the minimum growth rate (blue) is only slightly less than the rate at which the population mean evolves, owing to the more gradual number density increase over this period. However, this slight decrease brings the required size evolution closer to the merger model. We conclude that mergers are roughly consistent with producing the more modest size evolution at $z \lesssim 1$, assuming the same favorable choices of $\alpha$ and $\tau_e$. 

Broadly speaking, our more elaborate model reaches a similar conclusion to that we inferred from a na\"{i}ve consideration of the mean sizes at various redshifts (Section 5.3, Figure 10). However, even this refined model involves some questionable assumptions. First, we have neglected the contribution that measurement errors make to the width of the observed distribution, on the grounds that they are expected to make a small contribution. In Appendix B we discuss how our results would be impacted if the true measurement errors increase rapidly with redshift. Second, we have assumed that the descendants of quiescent galaxies are also quiescent, but some systems may be rejuvenated by secondary episodes of star formation \citep[e.g.,][]{Treu05}. Since our results are driven by the abundance of the most compact systems, which are overwhelmingly quiescent, we expect this to be a small effect. For example, only $\simeq 15\%$ of $z>1$ galaxies that are more compact than the median quiescent galaxy at the same redshift are classified as star-forming. Third, the lower-$z$ CF in our comparisons applies to a constant mass threshold of $\log M_* > 10.7$. Since we expect the population in place at high-$z$ to be continually growing in mass, an evolving mass threshold would be more appropriate. For the specific assembly rate $\dot{M_*}/M_* \approx 0.03~\textrm{dex} / \tau_e$ expected from our pair analysis, this translates to reductions in number density of $\approx 10-20\%$ over the redshift intervals we considered. Since the last two effects are modest and oppose one another, neglecting them is justified.

Finally, it is important to consider the stochasticity of the merger progress.  Obviously, every galaxy cannot undergo exactly 0.16 mergers with mass ratios $\mu_* > 0.1$ per timescale. In reality, since the expected number of mergers per timescale is significantly less than unity, many galaxies will experience \emph{no} such mergers over an interval of several Gyr. This retards the movement of the compact end of the distribution in Figure~\ref{fig:demo}a, leaving even more late compact remnants. Accounting for stochasticity would therefore only strengthen our conclusion that additional mechanisms are necessary to explain the rate of size evolution at $z \gtrsim 1$. 

In summary, our models for size growth via minor merging can reasonably account only for that observed at $z \lesssim 1$. The faster growth rate at higher redshift remains difficult to explain via merging alone, even when one accounts for the rapid buildup of the quiescent population over the same period.

\section{Discussion and Conclusions}

Using high-quality near-infrared imaging from WFC3/IR taken as part of the CANDELS survey, in conjunction with other multi-wavelength data in the UKIRT Ultra Deep Field and GOODS-South fields, we have compiled a uniform sample of 935 galaxies with stellar masses greater than 10$^{10.7} \msol$ and photometric redshifts $0.4<z<2.5$. Within this sample, the most compact objects at a given redshift are those with quiescent stellar populations. For this subsample, the mean half-light radius measured at fixed stellar mass grows by a factor of 3.5 over this interval. The growth rate per unit time is noticeably quicker at early cosmic epochs, corresponding to the redshift range $z \approx 1.3-2.5$.

We have explored the physical origin of this size growth in 404 quiescent galaxies over $0.4<z<2$ by searching for close pairs whose photometric redshifts imply a likely association with their hosts. The depth of the imaging allows us to probe secondary companions whose stellar masses are only 10\% of their primary hosts.  Our main conclusion is that the delivery of stellar mass in mergers, estimated via the incidence of close pairs, cannot account for more than roughly half of the minimum size growth that $z = 2$ quiescent galaxies must incur to avoid leaving a greater number of late compact remnants than is observed. At $z \lesssim 1$, on the other hand, mergers may account for most or all of the size growth rate, but only if a short merger timescale ($\sim 1$~Gyr) and fairly robust growth efficiency ($\alpha \sim 1.6$) are valid. These conclusions hold if the evolution of the mass-size relation is driven in part by the emergence of new, systematically larger quiescent galaxies. If this is not the case, then the merger rate will fall further short of that needed to drive the observed size growth.

Given the variety of theoretical and observational ingredients in this analysis, it is worthwhile to review the assumptions underlying this conclusion. Foremost is the uncertainty in the merger timescale and growth efficiency. Most of the results in Section 5 assume optimistic values for the theoretical parameters ($\tau_e = 1$~Gyr, $\alpha = 1.6$). Furthermore, all mergers in our models are dry and thus provide the maximum amount of size growth, whereas many minor mergers at high-redshift may in fact involve gas-rich secondaries (Section 4.4). We also note that our correction for unbound projected pairs ($C_{\textrm{mg}}$, Section 5.1) is not specifically calibrated for red galaxies, which are more strongly clustered, and may therefore understate this correction and thus overstate the merger rate. Altogether, it is therefore easy to argue that mergers produce less size growth than we have presented, but it is hard to see how the effect of mergers could be much larger. 

The tension at high redshift can be viewed as a consequence of the observation that the rate of size growth per unit time is considerably larger beyond $z \simeq1.3$, whereas the pair fraction remains nearly constant. One conceivable explanation is that the merger timescale declines with increasing redshift. However, current theoretical studies do not support this suggestion \citep{Kitzbichler08,Lotz11}. Incompleteness due to photometric redshift errors is a concern as higher redshifts are probed, but our best estimates of the catastrophic error rate (Section \ref{sec:photoz}) are not high enough to significantly alter our conclusions. We note also that although the energy arguments discussed in Section~\ref{sec:alpha} are generally applicable, the details of our framework for analyzing size growth are premised on spheroid-spheroid mergers. This is true for most other observational studies to date, since the theoretical framework for such mergers has been most extensively developed. Further studies of simulated spheroid-disk minor mergers, particularly with progenitors consistent with $z \simeq 2$ observations, are needed to better assess the growth efficiency when the incoming stellar material is more loosely bound. Still, our pair fraction measurements imply that only about $50~[\tau_e / 1~\textrm{Gyr}]^{-1}$ percent of $z \simeq 2$ quiescent galaxies experience any $\mu_* > 0.1$ mergers over $z = 1-2$. This is likely to pose a challenge regardless of the particular merger physics.
 
An equally important assumption is that the observed half-light radii are valid proxies for half-mass radii. The former are measured observationally, but the latter are relevant when considering the mass-structural changes caused by mergers. Although a detailed study of color gradients and their evolution in our CANDELS sample is beyond the scope of this paper, these data do confirm earlier studies that quiescent galaxies at $z \sim 2$ typically display negative color gradients (i.e., are bluer on the outside), and that these tend to flatten at lower redshift \citep{vanDokkum10,Guo11,Cassata11}. The color gradients probably arise from a complex combination of age, dust, and metallicity gradients, but in any case the stellar mass-to-light ratio is lower on the outside, so that these galaxies are \emph{more} compact in mass than in light. If anything, we therefore expect to have underestimated the rate of structural change.
 
Much of the early skepticism regarding the rapid size evolution of early-type galaxies focused on the possibility of severe observational errors in measuring the key parameters of size and mass. Stellar masses could be overestimated by imperfect population synthesis models, or effective radii could be underestimated in shallow imaging \citep[e.g.,][]{Mancini10}. Subsequent observations have weakened these claims. Although substantial uncertainties remain in stellar population synthesis models \citep{Muzzin09}, dynamical masses measured from absorption spectra in moderate samples at $z \sim 1.3$ \citep{Newman10} and for a few individual galaxies or stacked spectra at $z \sim 1.6-1.8$ \citep{Cappellari09,vandeSande11} have not indicated large systematic discrepancies with photometrically-determined stellar masses. Regarding size measurements, the CANDELS survey represents a major advance as it provides the first large space-based sample taking advantage of the improved depth and sampling of WFC3 relative to NICMOS. The radial surface brightness profile of a typical $z \sim 2$ quiescent galaxy in our sample can be traced to $\simeq 7R_e$.
 
Several theorists have compared the rate of galaxy size evolution in simulations to observations. \citet{Hopkins10b}, based on a suite of cosmological, hydrodynamical simulations, also conclude that mergers alone do not generate the entire rate of growth observed for quiescent galaxies. To explain the remainder, they propose a combination of several physical and observational effects. First, they assume that stellar masses and effective radii are over- and underestimated, respectively. Second, they suggest that the presence of blue cores implies that half-mass radii are larger than the measured half-light radii; as discussed above, the opposite appears more likely. Third, \citet{Hopkins10b} model adiabatic expansion due to mass loss from stellar winds, but this effect alone expands galaxies by only $\simeq 20\%$.  

\citet{Oser11}, on the other hand, present hydrodynamical ``zoom'' simulations in which galaxy size evolution at $z \lesssim 2$ agrees well with their compilation of observations and attribute the size expansion primarily to minor mergers. As these authors note, one concern is that the absence of supernova feedback in this set of simulations enhances the stellar mass formed in low-mass halos. This could overstate the effectiveness of minor mergers by substantially increasing the stellar mass they deliver. As simulations and observations at $z \simeq 2$ improve, it may be possible to test such effects through additional comparisons, such as the stellar mass--halo mass relation or the evolution of the stellar mass function.

C.~Nipoti et al (2011, submitted) construct a $\Lambda$CDM-based analytic framework, supported by suites of $N$-body spheroid merger simulations, to predict the evolution of early-type galaxies undergoing dry mergers. Using a compilation of observations of early-type galaxies at $z = 1 - 2.5$ (including this work), they conclude that mergers alone are not consistent with the observed rate of structural evolution at $z \gtrsim 1.3$. Following on earlier work \citep{Nipoti09}, they also find that mergers introduce too much scatter in the scaling relations at lower redshift unless the progenitors are finely tuned to occupy a very tight region in the mass--radius plane. Such fine tuning is not consistent with the near constancy of the scatter that we observe in this plane.

Future work can extend this study in many ways. Imaging of the remaining CANDELS fields will allow possible trends of sizes and pair fractions with mass, redshift, and environment to be discerned more clearly, which may shed light on the responsible physical mechanisms. Multiplexed near-infrared spectrographs soon to be commissioned on $8-10$~m telescopes will provide redshifts and confirmation of the quiescent nature for larger samples at high redshift than has previously been possible. This will provide an invaluable test of the photometric redshift and star formation rate estimates on which the present study depends, although with current telescopes we are likely to continue to rely on photometric estimates for many of the faint companions. It should also be possible to significantly enlarge the library of dynamical mass estimates, of which only a handful are currently available for quiescent galaxies at $z > 1.5$, and thereby test the accuracy and precision of stellar mass estimates at higher redshifts. Spectroscopic indicators of maturity and recent star formation activity (Balmer lines, 4000~\AA~break) may allow tests of the ``minimum growth'' hypothesis considered in this work, i.e., that early quiescent galaxies remain the most compact systems in place at later epochs. If this is not the case, the challenge of accounting physically for the rapid growth of quiescent galaxies will be further heightened.

\acknowledgments
We thank Carlo Nipoti and Anna Nierenberg for fruitful conversations and useful comments, as well as the anonymous referee for a helpful report. This work is based on observations taken by the CANDELS Multi-Cycle Treasury Program with the NASA/ESA \emph{HST}, which is operated by the Association of Universities for Research in Astronomy, Inc., under NASA contract NAS5-26555. T.T.~thanks the Packard Foundation for their support through a Packard Fellowship.

\bibliographystyle{apj}
\bibliography{newm1011}{}

\pagebreak
\clearpage
\appendix

\section{The Skew Normal Distribution}
In Section 3 we fit the distribution of sizes of quiescent galaxies to skew normal distributions that evolve with redshift in order to assess changes in the mean and dispersion. The skew normal distribution has the probability density function
\begin{equation}
P(x) = \frac{1}{\omega \pi} e^{-\frac{(x - \psi)^2}{2\omega^2}} \int_{-\infty}^{s \left(\frac{x-\psi}{\omega}\right)} e^{-\frac{t^2}{2}} dt,
\end{equation}
characterized by the parameters $(\psi, \omega, s)$. Throughput this paper, we use a parameterization in terms of the mean $\bar{x}$ and standard deviation $\sigma$, which relate to $(\psi,\omega,s)$ through the relations $\bar{x} = \psi + \omega \delta \sqrt{2/\pi}$ and $\sigma^2 = \omega^2 (1 - 2\delta^2/\pi)$, where $\delta = s / \sqrt{1+s^2}$. The shape parameter $s$ relates to the skewness, and $s = 0$ recovers a Gaussian distribution.

\section{Measurement Errors in Stellar Masses and Radii}
\begin{figure}
\centering
\includegraphics[width=0.9\linewidth]{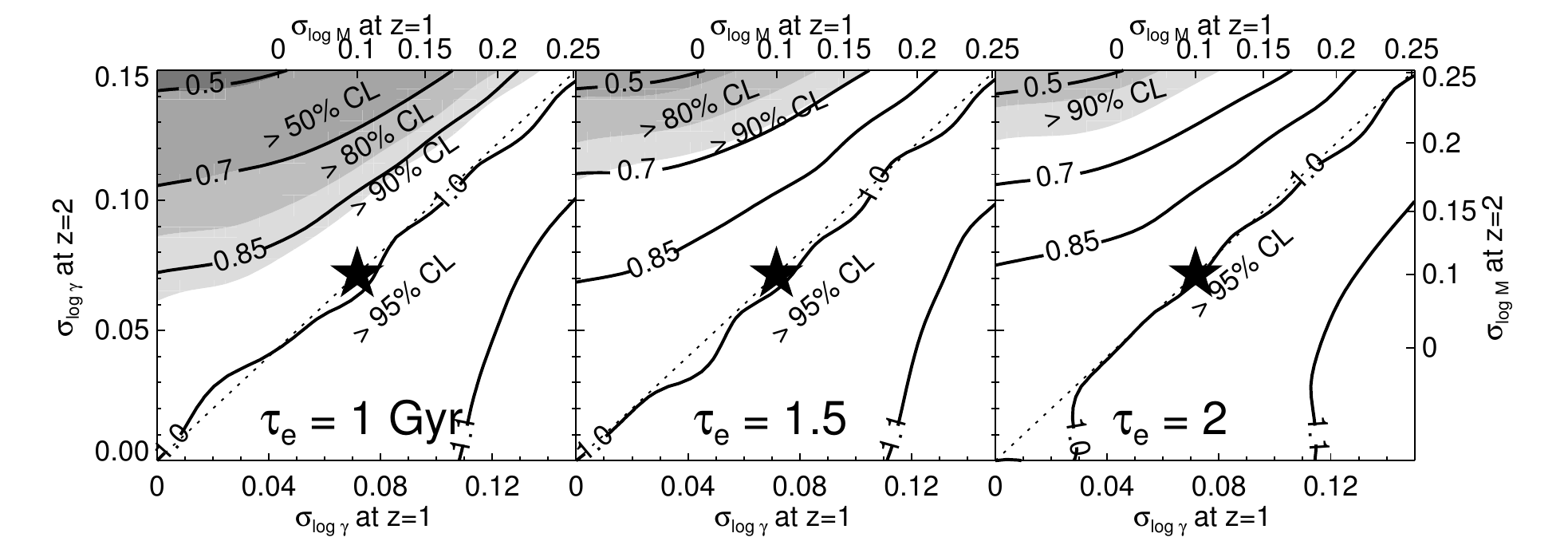}
\caption{Effects of measurement errors in stellar masses and radii on our conclusions. Contours show the factor by which $\Delta \log \gamma_{\textrm{min}}$, the minimum necessary growth for quiescent systems over $z = 1-2$, would change under different estimates of the measurement errors $\sigma_{\log \gamma}$ at $z = 1$ and 2. Top and right axes show the corresponding errors $\sigma_{\log M_*}$ in stellar mass, assuming 10\% errors in radii. Shaded regions indicate the corresponding confidence level (CL) at which merging alone as a viable growth mechanism is rejected. The star indicates the estimated errors based on the rms formal stellar mass uncertainties from SED fitting. Panels display results for several effective merger timescales $\tau_e$ and $\alpha = 1.6$. As discussed in Appendix B, if the uncertainties on stellar mass (modulo IMF choice) are the same (dotted line) at $z = 1$ and $z = 2$, then our results are unchanged, and merging alone as a growth mechanism is rejected at $>95\%$ CL. Only if the errors are much larger at $z = 2$ and the merger timescale is short is the CL reduced.\label{fig:widths}}
\end{figure}

As discussed in Sections 3 and 5, the distributions in $\gamma$ that we fit and compare to merger models are dominated by the intrinsic variation in $\gamma$, but also include some component of scatter arising from measurement errors in the radii and stellar masses. Random errors in stellar mass estimates are small with good photometry (typically $\sim 0.1$~dex in this work; see also \citealt{Auger09}), but systematic errors are not well understood. Comparison with independent dynamical mass measurements can place upper limits on the true scatter in stellar mass estimates. At $z \sim 0$, this limits the scatter to $\sigma_{\log M_*} \lesssim 0.15$~dex, based on the SDSS sample described in Section 2.6, while at $z \simeq 1.3$ the sample of spheroids from \citet{Newman10} indicates a similar scatter of $\sigma_{\log M_*} \lesssim 0.1-0.2$~dex when spectroscopic redshifts are used. Analogous comparisons are currently not possible at higher redshift. For this study, as we describe below, the absolute uncertainties are not as important as how they may evolve with redshift.

The main sources of systematic uncertainty include the unknown IMF and the complexities of stellar population synthesis models. The former is less critical for our analysis, since in our ``minimum growth'' test (Section 5.4) we are tracking the same sample of massive, quiescent galaxies, so the IMF should not change. The latter uncertainty is likely more important, since younger populations may be systematically different \citep[e.g.,][]{Maraston05,Conroy09}. A simple estimate of this effect can be obtained by comparing stellar mass estimates from the BC03 and CB07 models, which differ in their treatment of the TP-AGB stars that may dominate the NIR light at ages of $\sim 1$~Gyr. As discussed in Section 3, these models predict stellar masses systematically offset by $-0.05z$ for our quiescent sample. However, the scatter between the two models actually declines slightly from $\sim0.1$~dex at $z = 1$ to $\sim0.05$~dex at $z = 2$. 

The main concern for our ``minimum growth rate'' study is the reverse: that the scatter in stellar mass measurements for quiescent galaxies at $z = 2$ is much larger than at $z = 1$. In this scenario, the true abundance of very compact $z = 2$ galaxies would be smaller than our fits indicate, since some are simply scattered to small $\gamma$ through random errors in $R_h$ and $\log M_*$. If the $z = 1$ measurement errors are comparable, then this effect approximately cancels in our comparison of compactness functions at the two redshifts. But if the $z = 1$ measurement errors are smaller than those at $z = 2$, we expect that the minimum necessary size growth over $z = 1-2$ would lessen.

We can test the effects of redshift-dependent errors in our comparison of two compactness functions $\textrm{CF}_1$ and $\textrm{CF}_2$ with estimated measurement errors $\sigma_{\log \gamma, 1} > \sigma_{\log \gamma,2}$ by convolving $\textrm{CF}_2$ by a Gaussian with dispersion $\sigma = \sqrt{\sigma_{\log \gamma, 1}^2  - \sigma_{\log \gamma, 2}^2}$. Having thus matched the measurement errors, we then derive the minimum necessary growth $\Delta \log \gamma_{\textrm{min}}$ following Section~5.4. The contours in Figure~\ref{fig:widths} show how this minimum growth would change for evolution over $z = 1 - 2$ assuming various measurement errors at $z = 1$ and 2. The three panels consider a range of merger timescales $\tau_e$ spanning the range discussed in Section 5.1 and a growth efficiency of $\alpha = 1.6$. As anticipated, when the measurement errors are nearly equal (dotted line), the derived minimum growth is not affected. When $\sigma_{\log \gamma}$ is much greater at $z = 2$ than at $z = 1$, however, the true minimum growth rate may be smaller and thus more comparable to the rate attainable through merging. 

The shaded regions in Figure~\ref{fig:widths} display the confidence levels at which merging alone as a driver of size evolution is rejected. These panels indicate that the claim that merging alone is insufficient at $z \gtrsim 1$ is seriously weakened only if the measurement errors are significantly larger at $z = 2$ than at $z = 1$ \emph{and} the effective timescale $\tau_e$ is very short ($\simeq 1$ Gyr). Given the other assumptions entering this exercise that are favorable for mergers (namely, that \emph{all} of the most compact systems at $z = 1$ are descended from $z = 2$ quiescent galaxies, that all mergers are dry, and that there is no stochasticity in the incidence of mergers), we believe that our main results are robust.

\end{document}